\documentclass[12pt]{article}
\usepackage{amsmath}

\usepackage{amssymb}
\usepackage{bm}
\usepackage[dvips]{graphicx}

\newcommand{\llangle}{\langle\!\langle}
\newcommand{\rrangle}{\rangle\!\rangle}

\begin{document}

\baselineskip=17pt

\begin{titlepage}
\rightline{\tt arXiv:1508.00366}
\rightline{\tt UT-Komaba/15-4}
\rightline{\tt YITP-15-65}
\begin{center}
\vskip 2.5cm
{\Large \bf {Complete action for open superstring field theory}}
\vskip 1.0cm
{\large {Hiroshi Kunitomo${}^1$ and Yuji Okawa${}^2$}}
\vskip 1.0cm
${}^1${\it {Yukawa Institute for Theoretical Physics, Kyoto University}}\\
{\it {Kitashirakawa Oiwakecho, Sakyo-ku, Kyoto 606-8502, Japan}}\\
kunitomo@yukawa.kyoto-u.ac.jp
\vskip 0.5cm
${}^2${\it {Institute of Physics, The University of Tokyo}}\\
{\it {3-8-1 Komaba, Meguro-ku, Tokyo 153-8902, Japan}}\\
okawa@hep1.c.u-tokyo.ac.jp

\vskip 2.0cm

{\bf Abstract}
\end{center}

\noindent
We construct a complete action for open superstring field theory
that includes the Neveu-Schwarz sector and the Ramond sector.
For the Neveu-Schwarz sector, 
we use the string field in the large Hilbert space of the superconformal ghost sector,
and the action in the Neveu-Schwarz sector is the same
as the Wess-Zumino-Witten-like action of the Berkovits formulation.
For the Ramond sector,
it is known that the BRST cohomology on an appropriate subspace
of the small Hilbert space reproduces the correct spectrum,
and we use the string field projected to this subspace.
We show that the action is invariant under gauge transformations
that are consistent with the projection for the string field in the Ramond sector.

\end{titlepage}

\tableofcontents

\newpage

\section{Introduction}\label{Introduction}
\setcounter{equation}{0}

The open superstring in the Ramond-Neveu-Schwarz formalism
consists of the Neveu-Schwarz (NS) sector and the Ramond sector,
and a complete formulation of open superstring field theory requires
the inclusion of string fields of both sectors.
The central issue in formulating open superstring field theory
has been how we should tame the picture of open superstring fields.

For the NS sector, Berkovits constructed a Wess-Zumino-Witten-like (WZW-like) action~\cite{Berkovits:1995ab}
based on the large Hilbert space of the superconformal ghost sector~\cite{Friedan:1985ge}.
The open superstring field is in the $0$ picture,
and no picture-changing operators are used in the action.
Recently, it was demonstrated that a regular formulation
based on the small Hilbert space of the superconformal ghost sector
can be obtained from the Berkovits formulation by partial gauge fixing~\cite{Iimori:2013kha},
and then an action
with an $A_\infty$ structure~\cite{Stasheff:I, Stasheff:II, Getzler-Jones, Markl, Penkava:1994mu, Gaberdiel:1997ia}
was constructed in~\cite{Erler:2013xta}.\footnote{
The construction was further generalized to the NS sector
of heterotic string field theory and the NS-NS sector of type II superstring field theory
in~\cite{Erler:2014eba}.
See~\cite{Jurco:2013qra, Matsunaga:2013mba, Matsunaga:2014wpa, Goto:2015hpa}
for recent discussions on closed superstring field theory.
}
This is an important achievement because the $A_\infty$ structure plays a crucial role
when we quantize open superstring field theory
based on the Batalin-Vilkovisky formalism~\cite{Batalin:1981jr, Batalin:1984jr}.
When we explicitly construct interaction terms by carrying out the program of~\cite{Erler:2013xta},
however, the number of terms grows as we go to higher orders
and the form of the interactions will be extremely complicated.
On the other hand, the action of the Berkovits formulation
is beautifully written in the WZW-like form, and we have much better control
over the interaction terms,
although the WZW-like action does not exhibit the $A_\infty$ structure
and its Batalin-Vilkovisky quantization~\cite{Kroyter:2012ni, Torii:2012nj, Torii:2011zz, Berkovits:2012np, 
Iimori:2015aea}
has turned out to be formidably complicated~\cite{Berkovits:201X}.
Very recently, it was shown that the theory with the $A_\infty$ structure in~\cite{Erler:2013xta}
is related to the Berkovits formulation
by partial gauge fixing and field redefinition~\cite{Erler:2015rra, Erler:2015uba},
and we can now
extract the $A_\infty$ structure from the Berkovits formulation
by the field redefinition.

Inclusion of the string field in the Ramond sector was less successful,
and we did not have satisfactory formulations.
In the earlier approach in~\cite{Witten:1986qs}
or its modification~\cite{Preitschopf:1989fc, Arefeva:1989cp},
the string field of picture number $-1/2$
in the small Hilbert space was used.
For incorporation of the Ramond sector into the Berkovits formulation
based on the large Hilbert space,
the equations of motion were written in a covariant form~\cite{Berkovits:2001im},
but the action constructed in~\cite{Berkovits:2001im}
was not completely covariant,
although it respects the covariance for a class of interesting backgrounds
such as D3-branes in the flat 10D spacetime.
Another approach is to use a constraint to be imposed
on the equations of motion after they are derived from an action~\cite{Michishita:2004by}
as in type IIB supergravity.\footnote{
The Berkovits formulation of open superstring field theory
based on the large Hilbert space was extended
to the NS sector of heterotic string field theory~\cite{Okawa:2004ii, Berkovits:2004xh}.
The equations of motion including the Ramond sector
for heterotic string field theory were constructed in~\cite{Kunitomo:2013mqa, Kunitomo:2014hba},
and the approach in~\cite{Michishita:2004by}
was also extended to heterotic string field theory in~\cite{Kunitomo:2013mqa}.
While four-point amplitudes of the open superstring
including the Ramond states at the tree level
were correctly reproduced by the Feynman rules in~\cite{Michishita:2004by},
it was reported that correct five-point amplitudes were not reproduced~\cite{Michishita:unpublished}.
This issue was recently resolved in~\cite{Kunitomo:2014qla}
by correcting the Feynman rules; it was further extended to the action with a constraint
for heterotic string field theory~\cite{Kunitomo:2013mqa}
and correct four-point and five-point amplitudes
including the Ramond states at the tree level were reproduced~\cite{Kunitomo:2015hda}.
}
For the recent development of open superstring field theory
with the $A_\infty$ structure 
based on the small Hilbert space~\cite{Erler:2013xta},
the equations of motion including the Ramond sector were constructed in term of multi-string products
satisfying the $A_\infty$ relations~\cite{Erler:2015lya},
but an action to yield the equations of motion
including the Ramond sector
has not been constructed.

So what is the difficulty in constructing an action
including the string field in the Ramond sector?
The fundamental difficulty lies in the construction of the kinetic term
for the string field in the Ramond sector.
We consider that the source of the difficulty is related to the fact
that the propagator strip has a fermionic modulus in addition to the bosonic modulus
corresponding to the length of the strip when we regard propagator strips
as super-Riemann surfaces.
Let us explain this by comparing it with the open bosonic string
and the closed bosonic string.

The propagator strip in the open bosonic string
can be generated by the Virasoro generator $L_0$ as $e^{-t L_0}$,
and the parameter $t$ is the modulus corresponding to the length of the strip. 
In open bosonic string field theory~\cite{Witten:1985cc}, the integration
over this modulus is implemented by the propagator in Siegel gauge as
\begin{equation}
\frac{b_0}{L_0} = \int_0^\infty dt \, b_0 \, e^{-t L_0} \,,
\end{equation}
where the zero mode of the $b$ ghost $b_0$
is the ghost insertion associated with the integration
over this modulus.

The propagator surface in the closed bosonic string
can be generated by the Virasoro generators
$L_0 +\widetilde{L}_0$ and $i \, ( L_0 -\widetilde{L}_0 )$
as $e^{-t ( L_0 +\widetilde{L}_0 ) +i \theta ( L_0 -\widetilde{L}_0 )}$,
where $t$ and $\theta$ are moduli.
In closed bosonic string field theory,
whose construction~\cite{Kaku:1988zv, Kaku:1988zw, Saadi:1989tb, Kugo:1989aa, Kugo:1989tk}
was completed by Zwiebach in~\cite{Zwiebach:1992ie},
the integration over $t$ is implemented by the propagator in Siegel gauge
as in the open bosonic string:
\begin{equation}
\frac{b_0^+}{L_0^+}
= \int_0^\infty dt \, b_0^+ \, e^{-t \, L_0^+} \,,
\end{equation}
where
\begin{equation}
L_0^+ = L_0 +\widetilde{L}_0 \,, \qquad
b_0^+ = b_0 +\widetilde{b}_0 \,,
\end{equation}
and the sum of the zero modes $b_0$ and $\widetilde{b}_0$
of the holomorphic and antiholomorphic $b$ ghosts, respectively,
is the ghost insertion associated with the integration over the modulus $t$.
On the other hand, the integration over $\theta$ is implemented
as a constraint on the space of string fields.
The integration over $\theta$ yields the operator given by
\begin{equation}
B = b_0^- \, \int_0^{2 \pi} \frac{d \theta}{2 \pi} \, e^{i \theta L_0^-} \,,
\end{equation}
where
\begin{equation}
L_0^- = L_0 -\widetilde{L}_0 \,, \qquad
b_0^- = b_0 -\widetilde{b}_0 \,,
\end{equation}
and $b_0^-$ is the ghost insertion associated with the integration over this modulus.
The operator $B$ can be schematically understood as $\delta (b_0^-) \, \delta (L_0^-) \,$.
The closed bosonic string field $\Psi$ of ghost number $2$ is constrained to satisfy
\begin{equation}
b_0^- \, \Psi = 0 \,, \qquad L_0^- \, \Psi = 0 \,,
\end{equation}
and the BRST cohomology on this restricted space is known to give
the correct spectrum of the closed bosonic string.
The appropriate inner product of $\Psi_1$ and $\Psi_2$
satisfying the constraints can be written as the BPZ inner product
with an insertion of $c_0^-$ in the form
\begin{equation}
\langle \, \Psi_1, c_0^- \Psi_2 \, \rangle \,,
\end{equation}
where
$c_0^-$ consists of the zero modes $c_0$ and $\widetilde{c}_0$
of the holomorphic and antiholomorphic $c$ ghosts, respectively, as
\begin{equation}
c_0^- = \frac{1}{2} \, ( \, c_0 -\widetilde{c}_0 \, ) \,.
\end{equation}
The kinetic term of closed bosonic string field theory is then given by
\begin{equation}
S = {}-\frac{1}{2} \, \langle \, \Psi, c_0^- Q \Psi \, \rangle \,,
\end{equation}
where $Q$ is the BRST operator.
The operator $B$ can also be written as
\begin{equation}
B = -i \int_0^{2 \pi} \frac{d \theta}{2 \pi} \int d \tilde{\theta} \,
e^{i \theta L_0^- +i \tilde{\theta} \, b_0^-} \,,
\label{B}
\end{equation}
where $\tilde{\theta}$ is a Grassmann-odd variable,
and the extended BRST transformation introduced in~\cite{Witten:2012bh}
maps $\theta$ to $\tilde{\theta}$.
The extended BRST transformation acts in the same way
as the ordinary BRST transformation for operators
in the boundary conformal field theory (CFT),
and in particular it maps $b_0^-$ to $L_0^-$.
Therefore, the combination $i \theta L_0^- +i \tilde{\theta} \, b_0^-$
in~\eqref{B} is obtained from $i \theta \, b_0^-$
by the extended BRST transformation.
Note that the closed bosonic string field $\Psi$ satisfying the constraints
can be characterized as
\begin{equation}
B \, c_0^- \, \Psi = \Psi \,.
\label{closed-bosoinc-string-field-charaicterized}
\end{equation}

Let us now consider propagator strips
for the Ramond sector of the open superstring.
The fermionic direction of the moduli space can be parameterized as
$e^{\, \zeta G_0}$,  where $G_0$ is the zero mode of the supercurrent
and $\zeta$ is the fermionic modulus.
The integration over $\zeta$ with the associated ghost insertion
yields the operator $X$ given by
\begin{equation}
X = \int d \zeta \int d \tilde{\zeta} \, e^{\, \zeta G_0 -\tilde{\zeta} \, \beta_0} \,,
\label{X}
\end{equation}
where $\tilde{\zeta}$ is a Grassmann-even variable
and $\beta_0$ is the zero mode of the $\beta$ ghost.
The extended BRST transformation introduced in~\cite{Witten:2012bh}
maps $\zeta$ to $\tilde{\zeta}$
and maps $\beta_0$ to $G_0$
so that the combination $\zeta G_0 -\tilde{\zeta} \, \beta_0$
in~\eqref{X} is obtained from $-\zeta \beta_0$
by the extended BRST transformation.
If we perform the integration over $\zeta$, we obtain
\begin{equation}
X = {}-\delta ( \beta_0 ) \, G_0 +\delta' ( \beta_0 ) \, b_0 \,.
\label{PCO}
\end{equation}
See Appendix~\ref{appendix-X} for details.
It is known that the correct spectrum of the open superstring
can be reproduced by the BRST cohomology on the space
of open superstring fields for the Ramond sector of ghost number $1$
and picture number $-1/2$
that are restricted
to an appropriate form~\cite{Kazama:1985hd, Kazama:1986cy, Terao:1986ex}.
The appropriate inner product of $\Psi_1$ and $\Psi_2$ in the restricted space
can be written as the BPZ inner product in the small Hilbert space
with an insertion of $Y$ denoted by
\begin{equation}
\llangle \, \Psi_1, Y \Psi_2 \, \rrangle
\end{equation}
with
\begin{equation}
Y = {}-c_0 \, \delta' ( \gamma_0 ) \,,
\label{inversePCO}
\end{equation}
where $\gamma_0$ is the zero mode of the $\gamma$ ghost,
and the kinetic term of open superstring field theory for the Ramond sector is 
given by~\cite{Yamron:1986nb, Kazama:1985hd, Terao:1986ex}
\begin{equation}
S = {}-\frac{1}{2} \, \llangle \, \Psi, Y Q \Psi \, \rrangle \,.
\end{equation}
The important point is that the open superstring field $\Psi$ in the restricted space
can be characterized using the operator $X$~\eqref{PCO} as~\cite{Kugo:1988mf}
\begin{equation}
X Y \Psi = \Psi \,.
\end{equation}
This is analogous to~\eqref{closed-bosoinc-string-field-charaicterized}
for the closed bosonic string field,
and we regard this characterization of the string field in the Ramond sector as fundamental.

The next question is then whether we can introduce interactions
that are consistent with this restriction of the string field in the Ramond sector.
Recently, Sen constructed the equations of motion
of the one-particle irreducible effective superstring field theory
including the Ramond sector~\cite{Sen:2015hha}.
While the construction is for the heterotic string and the type II superstring,
the idea can be applied to the construction
of the classical equations of motion of open superstring field theory
including the Ramond sector.
A salient feature of the resulting equations of motion
is that the interaction terms of the equation of motion for the Ramond sector
are multiplied by a zero mode of the picture-changing operator.
The origin of the zero mode of the picture-changing operator
is the propagator in the Ramond sector,
and it is just a different way of integrating the fermionic modulus
of the propagator strip so that we can replace it
by the operator $X$ in~\eqref{PCO}.
Then the interaction terms of the equation of motion for the Ramond sector
are multiplied by $X$.
This is reminiscent of the equation of motion
of closed bosonic string field theory,
where the interaction terms of the equation of motion
are multiplied by $B$,
and this structure indicates
that the open superstring field for the Ramond sector
in the restricted space can be consistently used for the interacting theory.

In this paper, we construct a gauge-invariant action for open superstring field theory
including the NS sector and the Ramond sector.
We use the WZW-like action of the Berkovits formulation for the NS sector,
and we couple it to the open superstring field for the Ramond sector
in the restricted space.
This is the first construction of a complete action for open superstring field theory
in a covariant form.

The rest of the paper is organized as follows.
In Section~\ref{kinetic-term-section} we explain the kinetic terms
we use for the string field in the NS sector and for the string field in the Ramond sector.
In Section~\ref{cubic-quartic-section}
we construct cubic and quartic interactions
so that the action is invariant under nonlinearly extended gauge transformations
up to this order.
In Section~\ref{complete-action-section}
we present the complete action
and show that it is gauge invariant.
This is the main result of this paper.
In Section~\ref{Berkovits-section} we investigate the relation
between the equations of motion constructed by Berkovits in~\cite{Berkovits:2001im}
and ours.
Section~\ref{conclusions-discussion} is devoted to conclusions and discussion.

\section{Kinetic terms}\label{kinetic-term-section}
\setcounter{equation}{0}

An open superstring field is a state
in the boundary CFT
corresponding to the D-brane we are considering.
The boundary CFT consists of the matter sector, the $bc$ ghost sector,
and the superconformal ghost sector,
and the superconformal ghost sector can be described
either by $\beta (z)$ and $\gamma (z)$ or by $\xi (z)$, $\eta (z)$, and $\phi (z)$~\cite{Friedan:1985ge}.
The two descriptions are related as follows:
\begin{equation}
\beta (z) = \partial \xi (z) \, e^{-\phi (z)} \,, \qquad
\gamma (z) = e^{\phi (z)} \, \eta (z) \,.
\label{bosonization}
\end{equation}

The Hilbert space we usually use for the $\beta \gamma$ system
is smaller than the Hilbert space for $\xi (z)$, $\eta (z)$, and $\phi (z)$
and is called the {\it small Hilbert space}.
In the description in terms of $\xi (z)$, $\eta (z)$, and $\phi (z)$,
a state is in the small Hilbert space when it is annihilated
by the zero mode of $\eta (z)$.
We denote the zero mode of $\eta (z)$ by $\eta$,
and then the condition that a state $A$ is in the small Hilbert space
can be stated as
\begin{equation}
\eta A = 0 \,.
\end{equation}

The Hilbert space for $\xi (z)$, $\eta (z)$, and $\phi (z)$
is called the {\it large Hilbert space}.
Since the anticommutation relation of $\eta$ and the zero mode $\xi_0$
of $\xi (z)$ is
\begin{equation}
\{ \eta, \xi_0 \} = 1 \,,
\end{equation}
any state $\Phi$ in the large Hilbert space can be written as follows:
\begin{equation}
\Phi = \{ \eta, \xi_0 \} \, \Phi
= \eta \xi_0 \Phi +\xi_0 \eta \Phi
= \widetilde{\Phi} +\xi_0 \widehat{\Phi} \,,
\label{Phi-decomposition}
\end{equation}
where
\begin{equation}
\widetilde{\Phi} = \eta \xi_0 \Phi \,, \qquad \widehat{\Phi} = \eta \Phi \,.
\end{equation}
The states $\widetilde{\Phi}$ and $\widehat{\Phi}$ are in the small Hilbert space
because $\eta^2 = 0$.
Therefore, any state $\Phi$ in the large Hilbert space
can be decomposed into two states in the small Hilbert space this way.

For the NS sector, we use an open superstring field $\Phi$ in the large Hilbert space.
It is a Grassmann-even state, its ghost number is $0$, and its picture number is $0$.
The kinetic term $S_{NS}^{(0)}$ of $\Phi$ in the Berkovits formulation~\cite{Berkovits:1995ab}
is given by
\begin{equation}
S_{NS}^{(0)} = {}-\frac{1}{2} \, \langle \, \Phi, Q \eta \Phi \, \rangle \,,
\end{equation}
where $Q$ is the BRST operator and $\langle \, A, B \, \rangle$
is the BPZ inner product of $A$ and $B$.
The action is invariant under the gauge transformations given by
\begin{equation}
\delta_\Lambda^{(0)} \Phi = Q \Lambda \,, \qquad
\delta_\Omega^{(0)} \Phi = \eta \Omega \,,
\end{equation}
where $\Lambda$ and $\Omega$ are gauge parameters in the NS sector.
The gauge invariance can be shown by the following properties:
\begin{equation}
\begin{split}
& Q^2 = 0 \,, \qquad \eta^2 = 0 \,, \qquad \{ Q, \eta \} = 0 \,, \qquad
\langle \, B, A \, \rangle = (-1)^{AB} \langle \, A, B \, \rangle \,, \\
& \langle \, Q A, B \, \rangle = {}-(-1)^A \, \langle \, A, Q B \, \rangle \,, \qquad
\langle \, \eta A, B \, \rangle = {}-(-1)^A \, \langle \, A, \eta B \, \rangle \,.
\end{split}
\end{equation}
Here and in what follows, a state in the exponent of $-1$ represents
its Grassmann parity: it is $0$ mod $2$ for a Grassmann-even state
and $1$ mod $2$ for a Grassmann-odd state.

The equation of motion of the free theory is given by
\begin{equation}
Q \eta \Phi = 0 \,.
\end{equation}
As in~\eqref{Phi-decomposition},
we write $\Phi$ as $\Phi = \widetilde{\Phi} +\xi_0 \widehat{\Phi}$,
and we can bring $\Phi$ to the form $\xi_0 \widehat{\Phi}$
by the gauge transformation $\delta_\Omega^{(0)} \Phi$
with $\Omega = -\xi_0 \widetilde{\Phi}$.
Then the equation of motion reduces to the following form:
\begin{equation}
Q \eta \xi_0 \widehat{\Phi} = Q \, \{ \eta, \xi_0 \} \, \widehat{\Phi} = Q \widehat{\Phi} = 0 \,.
\end{equation}
The string field $\Phi$ brought to the form $\xi_0 \widehat{\Phi}$
satisfies the condition $\eta \xi_0 \Phi= 0$,
and the gauge transformation $\delta \Phi = Q \Lambda +\eta \Omega$
preserving this condition should satisfy
$\eta \xi_0 \, \delta \Phi = 0$. This constrains the gauge parameters as follows:
\begin{equation}
\eta \xi_0 \, \delta \Phi = \eta \xi_0 Q \Lambda +\eta \xi_0 \eta \Omega
= \eta \xi_0 Q \Lambda +\eta \Omega = 0 \,.
\end{equation}
We therefore choose $\eta \Omega$ to be $-\eta \xi_0 Q \Lambda$ and find
\begin{equation}
\delta \Phi = Q \Lambda -\eta \xi_0 Q \Lambda
= \xi_0 \eta Q \Lambda = {}-\xi_0 Q \eta \Lambda \,.
\end{equation}
This generates the transformation of $\widehat{\Phi}$ given by
\begin{equation}
\delta \widehat{\Phi} = \eta \delta \Phi
= {}-\eta \xi_0 Q \eta \Lambda = Q \widehat{\Lambda}
\end{equation}
with $\widehat{\Lambda} = -\eta \Lambda$ in the small Hilbert space.
This way the physical state condition $Q \widehat{\Phi} = 0$ in the small Hilbert space
and the equivalence relation
$\widehat{\Phi} \sim \widehat{\Phi} +Q \widehat{\Lambda}$
are reproduced.
This partial gauge fixing can be extended to the interacting theory.
See~\cite{Iimori:2013kha} for details.

For the Ramond sector, we use an open superstring field $\Psi$ in the small Hilbert space:
\begin{equation}
\eta \Psi = 0 \,.
\end{equation}
It is a Grassmann-odd state, its ghost number is $1$, and its picture number is $-1/2$.
We expand $\Psi$ based on the zero modes $b_0$, $c_0$, $\beta_0$, and $\gamma_0$ as
\begin{equation}
\Psi = \sum_{n=0}^\infty \, ( \gamma_0 )^n \, ( \phi_n +c_0 \, \psi_n ) \,, 
\end{equation}
where
\begin{equation}
b_0 \, \phi_n = 0 \,, \qquad \beta_0 \, \phi_n = 0 \,, \qquad
b_0 \, \psi_n = 0 \,, \qquad \beta_0 \, \psi_n = 0 \,.
\end{equation}
It is known~\cite{Kazama:1985hd, Kazama:1986cy, Terao:1986ex}
that the physical state condition can be written as
\begin{equation}
Q \Psi = 0
\end{equation}
with $\Psi$ restricted to the following form:
\begin{equation}
\Psi = \phi -( \gamma_0 +c_0 \, G \, ) \, \psi \,,
\end{equation}
where $G = G_0 +2 \, b_0 \gamma_0$ and
\begin{equation}
b_0 \, \phi = 0 \,, \qquad \beta_0 \, \phi = 0 \,, \qquad
b_0 \, \psi = 0 \,, \qquad \beta_0 \, \psi = 0 \,.
\end{equation}
See also~\cite{Yamron:1986nb, Ito:1985qa, FigueroaO'Farrill:1988hu}.
As pointed out in~\cite{Kugo:1988mf}, the string field $\Psi$ of this restricted form
can be characterized as
\begin{equation}
X Y \Psi = \Psi \,,
\label{Psi-condition}
\end{equation}
where the Grassmann-even operators $X$ and $Y$ are defined by\footnote{
The operators $\delta ( \beta_0 )$, $\delta' ( \beta_0 )$, and $\delta' ( \gamma_0 )$ here
and the operators $\delta' ( \gamma (z) )$ and $\Theta ( \beta_0 )$ that will appear later
are Grassmann odd,
and it should be understood that an appropriate Klein factor is included when it is necessary.
}
\begin{equation}
X = {}- \delta ( \beta_0 ) \, G_0 +\delta' ( \beta_0 ) \, b_0 \,, \qquad
Y = {}- c_0 \, \delta' ( \gamma_0 ) \,.
\end{equation}
The picture number of $X$ is $1$ and the picture number of $Y$ is $-1$.
As we mentioned in the introduction, the operator $X$
is related to the integration of the fermionic modulus
of propagator strips in the Ramond sector.
See Appendix~\ref{appendix-X} for details.\footnote{
For the geometric meaning of $X$ and $Y$,
see also~\cite{Belopolsky:1997jz}.
}
Therefore, the condition~\eqref{Psi-condition} on $\Psi$ can be understood
in the context of the supermoduli space of super-Riemann surfaces.
The operators $X$ and $Y$ satisfy the following relations:
\begin{equation}
X Y X = X \,, \qquad Y X Y = Y \,, \qquad [ \, Q, X \, ] = 0 \,, \qquad
\eta X \eta = 0 \,, \qquad \eta Y \eta = 0 \,.
\end{equation}
It then follows that the operator $X Y$ is a projector:
\begin{equation}
( X Y )^2 = X Y \,.
\end{equation}
We say that $\Psi$ is in the {\it restricted space}
when $\Psi$ satisfies
\begin{equation}
X Y \Psi = \Psi \,.
\end{equation}
While we always consider $\Psi$ of picture number $-1/2$,
we allow $\Psi$ to have an arbitrary ghost number
when we refer to the restricted space.
When $\Psi$ is in the restricted space,
$Q \Psi$ is also in the restricted space because
\begin{equation}
X Y Q \Psi = X Y Q X Y \Psi = X Y X Q Y \Psi = X Q Y \Psi = Q X Y \Psi = Q \Psi \,. 
\end{equation}
To summarize, the physical state condition and the equivalence relation can be stated as
\begin{equation}
Q \Psi = 0 \,, \qquad \Psi \sim \Psi +Q \lambda
\end{equation}
with $\Psi$ and $\lambda$ satisfying
\begin{equation}
\eta \Psi = 0 \,, \qquad X Y \Psi = \Psi \,, \qquad
\eta \lambda = 0 \,, \qquad X Y \lambda = \lambda \,.
\end{equation}

The appropriate inner product for $\Psi_1$ and $\Psi_2$ in the restricted space is
\begin{equation}
\llangle \, \Psi_1, Y \Psi_2 \, \rrangle \,,
\end{equation}
where $\llangle \, A, B \, \rrangle$ is the BPZ inner product of $A$ and $B$
in the small Hilbert space.
Recall that the picture number of $Y$ is $-1$,
and the total picture number is $-2$ for $\Psi_1$ and $\Psi_2$ of picture number $-1/2$. 
As we will show later, the operator $X$ is BPZ even in the small Hilbert space:
\begin{equation}
\llangle \, X A, B \, \rrangle = \llangle \, A, X B \, \rrangle \,.
\end{equation}
For $\Psi_1$ and $\Psi_2$ in the restricted space, we then have
\begin{align}
\llangle \, \Psi_1, Y \Psi_2 \, \rrangle
& = (-1)^{\Psi_1 \Psi_2} \, \llangle \, \Psi_2, Y \Psi_1 \, \rrangle \,,
\label{exchange-with-Y} \\
\llangle \, Q \Psi_1, Y \Psi_2 \, \rrangle
& = {}-(-1)^{\Psi_1} \, \llangle \, \Psi_1, Y Q \Psi_2 \, \rrangle \,.
\label{BPZ-Q-with-Y}
\end{align}
The relation~\eqref{exchange-with-Y} can be shown as
\begin{equation}
\begin{split}
\llangle \, \Psi_1, Y \Psi_2 \, \rrangle
& = \llangle \, X Y \Psi_1, Y \Psi_2 \, \rrangle
= \llangle \, Y \Psi_1, X Y \Psi_2 \, \rrangle \\
& = \llangle \, Y \Psi_1, \Psi_2 \, \rrangle
= (-1)^{\Psi_1 \Psi_2} \, \llangle \, \Psi_2, Y \Psi_1 \, \rrangle \,,
\end{split}
\end{equation}
and the relation~\eqref{BPZ-Q-with-Y} can be shown as
\begin{equation}
\begin{split}
\llangle \, Q \Psi_1, Y \Psi_2 \, \rrangle
& = \llangle \, Q X Y \Psi_1, Y \Psi_2 \, \rrangle
= \llangle \, Q Y \Psi_1, X Y \Psi_2 \, \rrangle \\
& = {}-(-1)^{\Psi_1} \, \llangle \, Y \Psi_1, Q X Y \Psi_2 \, \rrangle
= {}-(-1)^{\Psi_1} \, \llangle \, Y \Psi_1, X Y Q \Psi_2 \, \rrangle \\
& = {}-(-1)^{\Psi_1} \, \llangle \, X Y \Psi_1, Y Q \Psi_2 \, \rrangle
= {}-(-1)^{\Psi_1} \, \llangle \, \Psi_1, Y Q \Psi_2 \, \rrangle \,.
\end{split}
\end{equation}

We take the kinetic term $S_R^{(0)}$ for the Ramond sector
to be~\cite{Yamron:1986nb, Kazama:1985hd, Terao:1986ex}
\begin{equation}
S_R^{(0)} = {}-\frac{1}{2} \, \llangle \, \Psi, Y Q \Psi \, \rrangle
\end{equation}
for $\Psi$ satisfying
\begin{equation}
\eta \Psi = 0 \,, \qquad X Y \Psi = \Psi \,.
\end{equation}
The action is invariant under the gauge transformation
\begin{equation}
\delta_\lambda^{(0)} \Psi = Q \lambda \,,
\end{equation}
where $\lambda$ is a gauge parameter in the Ramond sector satisfying
\begin{equation}
\eta \lambda = 0 \,, \qquad X Y \lambda = \lambda \,.
\end{equation}
The equation of motion reproduces the physical state condition,
and the equivalence relation is implemented as a gauge symmetry.
The properties of the open superstring fields and the gauge parameters
are summarized in Table~\ref{table:string-fields}.
\begin{table}[ht]
\begin{center}
{\renewcommand\arraystretch{1.5}
\begin{tabular}{|c||c|c||c|c|c|}
\hline
field & $\Phi$ & $\Psi$ & $\Lambda$ & $\Omega$ & $\lambda$ \\
\hline
Grassmann & even & odd & odd & odd & even \\
\hline
$( \boldsymbol{g}, \boldsymbol{p} )$ & $(0,0)$ & $(1,-1/2)$ & $(-1,0)$ & $(-1,1)$ & $(0,-1/2)$ \\
\hline
\end{tabular}
}
\caption{Properties of the string fields and the gauge parameters.
The string field $\Phi$ in the NS sector is a Grassmann-even state,
and the string field $\Psi$ in the Ramond sector is a Grassmann-odd state.
The gauge parameters $\Lambda$ and $\Omega$ in the NS sector
are Grassmann-odd states,
and the gauge parameter $\lambda$ in the Ramond sector
is a Grassmann-even state.
The ghost number $\boldsymbol{g}$ and the picture number $\boldsymbol{p}$
of the string fields and the gauge fields are also shown.}
\label{table:string-fields}
\end{center}
\end{table}
The constraint on $\Psi$ characterized as $X Y \Psi = \Psi$ also plays a crucial role
in the context of the Batalin-Vilkovisky quantization~\cite{Kohriki:2012pp}.

The operator $Y$ in the kinetic term can be replaced by $Y_{\rm mid}$,
which is an insertion of $Y (z) = {}-c(z) \,  \delta' ( \gamma (z) )$
at the open-string midpoint:
\begin{equation}
{}-\frac{1}{2} \, \llangle \, \Psi, Y Q \Psi \, \rrangle
= {}-\frac{1}{2} \, \llangle \, \Psi, Y_{\rm mid} \, Q \Psi \, \rrangle \,.
\end{equation}
This can be shown from the relation $X Y_{\rm mid} \, X = X$ as follows:
\begin{equation}
\begin{split}
\llangle \, \Psi, Y Q \Psi \, \rrangle
& = \llangle \, X Y \Psi, Y X Y Q \Psi \, \rrangle
= \llangle \, Y \Psi, X Y X Y Q \Psi \, \rrangle
= \llangle \, Y \Psi, X Y Q \Psi \, \rrangle \\
& = \llangle \, Y \Psi, X Y_{\rm mid} \, X Y Q \Psi \, \rrangle
= \llangle \, X Y \Psi, Y_{\rm mid} \, X Y Q \Psi \, \rrangle
= \llangle \, \Psi, Y_{\rm mid} \, Q \Psi \, \rrangle \,.
\end{split}
\end{equation}
Therefore, our kinetic term coincides with that of open superstring field theory
in the Witten formulation~\cite{Witten:1986qs} for $\Psi$ in the restricted space.

We will construct interactions that couple $\Phi$ in the large Hilbert space
and $\Psi$ in the small Hilbert space.
Let us describe further the relation
between the large Hilbert space and the small Hilbert space.
The BPZ inner product $\llangle \, A, B \, \rrangle$
in the small Hilbert space defined for $A$ and $B$
satisfying $\eta A = 0$ and $\eta B = 0$
is related to the BPZ product in the large Hilbert space $\langle \, A, B \, \rangle$
as follows:
\begin{equation}
\llangle \, A, B \, \rrangle = \langle \, \xi_ 0 A, B \, \rangle \,.
\end{equation}
Since the zero mode $\xi_0$ is BPZ even, this can also be written as
\begin{equation}
\llangle \, A, B \, \rrangle = (-1)^A \, \langle \, A, \xi_ 0 B \, \rangle \,.
\end{equation}
The BRST cohomology is trivial in the large Hilbert space,
and thus the operator $X$, which commutes with the BRST operator, can be written as
\begin{equation}
X = \{ Q, \Xi \} \,,
\end{equation}
where $\Xi$ is a Grassmann-odd operator carrying ghost number $-1$ and picture number $1$.
We use $\Xi$ defined by~\cite{Preitschopf:1989fc}
\begin{equation}
\Xi = \Theta ( \beta_0 ) \,,
\end{equation}
where $\Theta$ is the Heaviside step function.
As we show in Appendix~\ref{appendix-Xi}, the anticommutator of $\eta$ and $\Xi$ is given by
\begin{equation}
\{ \eta, \Xi \} = 1 \,,
\label{eta-Xi}
\end{equation}
and $\Xi$ is BPZ even:
\begin{equation}
\langle \, \Xi A, B \, \rangle =  (-1)^A \langle \, A, \Xi B \, \rangle \,.
\label{BPZ-Xi}
\end{equation}
Because of the relation~\eqref{eta-Xi}
we can also use $\Xi$ to relate the BPZ inner product in the large Hilbert space
and the BPZ inner product in the small Hilbert space: 
\begin{equation}
\llangle \, A, B \, \rrangle = \langle \, \Xi A, B \, \rangle \,, \qquad
\llangle \, A, B \, \rrangle
= (-1)^A \, \langle \, A, \Xi B \, \rangle \,.
\end{equation}

Finally, let us discuss the BPZ property of the operator $X$.
Even when we work in the large Hilbert space, the operator $X$
always acts on a state in the small Hilbert space of picture number~$-3/2$,
and we show that $X$ is BPZ even in the small Hilbert space.
Actually, this can be shown even when $\Xi$ is not BPZ even,
and it follows only from the relation $\eta \Xi^\star +\Xi \eta = 1$
on a state of picture number~$-1/2$,
where $\Xi^\star$ is the BPZ conjugate of $\Xi$,
together with $\eta \Xi A = A$ and $\eta \Xi B = B$
for a pair of states $A$ and $B$ in the small Hilbert space of picture number~$-3/2$:
\begin{equation}
\begin{split}
\llangle \, X A, B \, \rrangle
& = (-1)^A \langle \, ( \, Q \Xi +\Xi Q \, ) \, A, \Xi B \, \rangle \\
& = (-1)^A \langle \, ( \, Q \Xi +\Xi Q \, ) \, \eta \Xi A, \Xi B \, \rangle
= (-1)^A \langle \, \eta \, ( \, Q \Xi^\star +\Xi^\star Q \, ) \, \Xi A, \Xi B \, \rangle \\
& = \langle \, ( \, Q \Xi^\star +\Xi^\star Q \, ) \, \Xi A, \eta \Xi B \, \rangle
= \langle \, \Xi A, ( \, \Xi Q +Q \Xi \, ) \, \eta \Xi B \, \rangle \\
& = \langle \, \Xi A, ( \, \Xi Q +Q \Xi \, ) \, B \, \rangle
= \llangle \, A, X B \, \rrangle \,.
\end{split}
\end{equation}

\section{Cubic and quartic interactions}\label{cubic-quartic-section}
\setcounter{equation}{0}

In this section we construct cubic and quartic terms of the action in the Ramond sector.
The action $S$ consists of $S_{NS}$ for the NS sector and $S_R$ for the Ramond sector:
\begin{equation}
S = S_{NS} +S_R \,,
\end{equation}
where $S_{NS}$ contains only $\Phi$
and $S_R$ contains both $\Phi$ and $\Psi$.
We expand $S_{NS}$ and $S_R$ as follows:
\begin{align}
S_{NS} & = S_{NS}^{(0)} +g \, S_{NS}^{(1)} +g^2 \, S_{NS}^{(2)} +O(g^3) \,, \\
S_R & = S_R^{(0)} +g \, S_R^{(1)} +g^2 \, S_R^{(2)} +O(g^3) \,,
\end{align}
where $g$ is the coupling constant and
\begin{align}
S_{NS}^{(0)} & = {}-\frac{1}{2} \, \langle \, \Phi, Q \eta \Phi \, \rangle \,, \\
S_R^{(0)} & = {}-\frac{1}{2} \, \llangle \, \Psi, Y Q \Psi \, \rrangle \,.
\end{align}
We also expand the gauge transformations as follows:
\begin{align}
\delta_\Lambda \Phi
& = \delta_\Lambda^{(0)} \Phi +g \, \delta_\Lambda^{(1)} \Phi +g^2 \, \delta_\Lambda^{(2)} \Phi +O(g^3) \,, \\
\delta_\Lambda \Psi
& = \delta_\Lambda^{(0)} \Psi +g \, \delta_\Lambda^{(1)} \Psi +g^2 \, \delta_\Lambda^{(2)} \Psi +O(g^3)
\end{align}
with
\begin{equation}
\delta_\Lambda^{(0)} \Phi = Q \Lambda \,, \qquad
\delta_\Lambda^{(0)} \Psi = 0 \,,
\end{equation}
where $\Lambda$ is a gauge parameter in the NS sector;
\begin{align}
\delta_\Omega \Phi
& = \delta_\Omega^{(0)} \Phi +g \, \delta_\Omega^{(1)} \Phi +g^2 \, \delta_\Omega^{(2)} \Phi +O(g^3) \,, \\
\delta_\Omega \Psi
& = \delta_\Omega^{(0)} \Psi +g \, \delta_\Omega^{(1)} \Psi +g^2 \, \delta_\Omega^{(2)} \Psi +O(g^3)
\end{align}
with
\begin{equation}
\delta_\Omega^{(0)} \Phi = \eta \Omega \,, \qquad
\delta_\Omega^{(0)} \Psi = 0 \,,
\end{equation}
where $\Omega$ is a gauge parameter in the NS sector;
and
\begin{align}
\delta_\lambda \Phi
& = \delta_\lambda^{(0)} \Phi +g \, \delta_\lambda^{(1)} \Phi +g^2 \, \delta_\lambda^{(2)} \Phi +O(g^3) \,, \\
\delta_\lambda \Psi
& = \delta_\lambda^{(0)} \Psi +g \, \delta_\lambda^{(1)} \Psi +g^2 \, \delta_\lambda^{(2)} \Psi +O(g^3)
\end{align}
with
\begin{equation}
\delta_\lambda^{(0)} \Phi = 0 \,, \qquad
\delta_\lambda^{(0)} \Psi = Q \lambda \,,
\end{equation}
where $\lambda$ is a gauge parameter in the Ramond sector.

For the NS sector, we use the cubic and quartic terms
in the Berkovits formulation~\cite{Berkovits:1995ab}:
\begin{align}
S_{NS}^{(1)} & = {}-\frac{1}{6} \, \langle \, \Phi, Q \, [ \, \Phi, \eta \Phi \, ] \, \rangle \,, \\
S_{NS}^{(2)} &
= {}-\frac{1}{24} \, \langle \, \Phi, Q \, [ \, \Phi, [ \, \Phi, \eta \Phi \, ] \, ] \, \rangle \,.
\end{align}
The gauge invariance up to this order can be stated as
\begin{align}
\delta_\Lambda^{(0)} S_{NS}^{(1)} +\delta_\Lambda^{(1)} S_{NS}^{(0)} = 0 \,, \qquad
\delta_\Lambda^{(0)} S_{NS}^{(2)} +\delta_\Lambda^{(1)} S_{NS}^{(1)}
+\delta_\Lambda^{(2NS)} S_{NS}^{(0)} = 0 \,,
\end{align}
where
\begin{align}
\delta_\Lambda^{(1)} \Phi
= {}-\frac{1}{2} \, [ \, \Phi, Q \Lambda \, ] \,, \qquad
\delta_\Lambda^{(2NS)} \Phi
= \frac{1}{12} \, [ \, \Phi, [ \, \Phi, Q \Lambda \, ] \, ] \,,
\end{align}
and
\begin{align}
\delta_\Omega^{(0)} S_{NS}^{(1)} +\delta_\Omega^{(1)} S_{NS}^{(0)} = 0 \,, \qquad
\delta_\Omega^{(0)} S_{NS}^{(2)} +\delta_\Omega^{(1)} S_{NS}^{(1)}
+\delta_\Omega^{(2)} S_{NS}^{(0)} = 0 \,,
\end{align}
where
\begin{align}
\delta_\Omega^{(1)} \Phi
= \frac{1}{2} \, [ \, \Phi, \eta \Omega \, ] \,, \qquad
\delta_\Omega^{(2)} \Phi
= \frac{1}{12} \, [ \, \Phi, [ \, \Phi, \eta \Omega \, ] \, ] \,.
\end{align}
As we will see, 
there is an additional contribution $\delta_\Lambda^{(2R)} \Phi$ to $\delta_\Lambda^{(2)} \Phi$
when we include the Ramond sector,
and $\delta_\Lambda^{(2)} \Phi$ is given by
\begin{equation}
\delta_\Lambda^{(2)} \Phi = \delta_\Lambda^{(2NS)} \Phi +\delta_\Lambda^{(2R)} \Phi \,.
\end{equation}
On the other hand, it will turn out that there are no corrections to $\delta_\Omega^{(2)} \Phi$
when we include the Ramond sector.
The goal of this section is to determine $S_R^{(1)}$ and $S_R^{(2)}$ in the action
and 
$\delta_\Lambda^{(1)} \Phi$,
$\delta_\Lambda^{(1)} \Psi$,
$\delta_\Omega^{(1)} \Phi$,
$\delta_\Omega^{(1)} \Psi$,
$\delta_\lambda^{(1)} \Phi$,
$\delta_\lambda^{(1)} \Psi$,
$\delta_\Lambda^{(2)} \Phi$,
$\delta_\Lambda^{(2)} \Psi$,
$\delta_\Omega^{(2)} \Phi$,
$\delta_\Omega^{(2)} \Psi$,
$\delta_\lambda^{(2)} \Phi$,
and $\delta_\lambda^{(2)} \Psi$ in the gauge transformations.

We use the star product~\cite{Witten:1985cc}
in constructing interaction terms,
and all products of string fields in this paper are defined by the star product.
The star product has the following properties:
\begin{equation}
\begin{split}
& ( \, A B \, ) \, C = A \, ( \, B C \, ) \,, \qquad
\langle \, A, B C \, \rangle = \langle \, A B, C \, \rangle \,, \\
& Q \, ( A B ) = ( Q A ) \, B +(-1)^A A \, ( Q B ) \,, \qquad
\eta \, ( A B ) = ( \eta A ) \, B +(-1)^A A \, ( \eta B ) \,. 
\end{split}
\end{equation}

We will construct cubic and quartic interactions
such that the action is invariant under nonlinearly extended gauge transformations.
Corrections to the gauge transformations are determined from the structures
of the kinetic terms in the following way.
The variation of $S_{NS}^{(0)}$ is given by
\begin{equation}
\delta S_{NS}^{(0)} = {}-\langle \, \delta \Phi, Q \eta \Phi \, \rangle \,.
\end{equation}
Therefore, a term of the form
\begin{equation}
\delta S = \langle \, A, Q \eta \Phi \, \rangle
\end{equation}
in the gauge variation can be canceled by $\delta S_{NS}^{(0)}$
with $\delta \Phi$ given by
\begin{equation}
\delta \Phi = A \,. 
\end{equation}
The variation of $S_R^{(0)}$ is given by
\begin{equation}
\delta S_R^{(0)} = {}-\llangle \, \delta \Psi, Y Q \Psi \, \rrangle \,.
\end{equation}
A term of the form
\begin{equation}
\delta S = \langle B, Q \Psi \, \rangle
\label{Q-Psi-term}
\end{equation}
in the gauge variation can be transformed as
\begin{equation}
\delta S = \langle \, B, \eta \, \xi_0 \, X Y Q \Psi \, \rangle
= \langle \, \xi_0 \eta B, X Y Q \Psi \, \rangle
= \llangle \, \eta B, X Y Q \Psi \, \rrangle
= \llangle \, X \eta B, Y Q \Psi \, \rrangle \,.
\end{equation}
Therefore, this can be canceled by $\delta S_R^{(0)}$
with $\delta \Psi$ given by
\begin{equation}
\delta \Psi = X \eta B \,. 
\end{equation}
Note that this form of $\delta \Psi$ satisfies the conditions
\begin{equation}
\eta \, \delta \Psi = 0 \,, \qquad X Y \delta \Psi = \delta \Psi \,.
\end{equation}

\subsection{The cubic interaction}

Let us consider the cubic interaction $S_R^{(1)}$ in the form
\begin{equation}
S_R^{(1)} = \alpha_1 \, \langle \, \Phi, \Psi^2 \, \rangle \,,
\end{equation}
where $\alpha_1$ is a constant to be determined.
When we take the string field $\Phi$ to be an on-shell state in the $-1$~picture multiplied by $\xi_0$
and the two string fields of $\Psi$ to be on-shell states in the $-1/2$~picture,
this cubic interaction reproduces
correct three-point amplitudes up to an overall normalization.
The action is gauge invariant at this order if we can find
$\delta_\Lambda^{(1)} \Psi$, $\delta_\Omega^{(1)} \Psi$,
$\delta_\lambda^{(1)} \Phi$, and $\delta_\lambda^{(1)} \Psi$
such that
\begin{equation}
\begin{split}
\delta_\Lambda^{(0)} S_R^{(1)} +\delta_\Lambda^{(1)} S_R^{(0)} & = 0 \,, \\
\delta_\Omega^{(0)} S_R^{(1)} +\delta_\Omega^{(1)} S_R^{(0)} & = 0 \,, \\
\delta_\lambda^{(0)} S_R^{(1)} +\delta_\lambda^{(1)} S_{NS}^{(0)} +\delta_\lambda^{(1)} S_R^{(0)} & = 0
\end{split}
\end{equation}
are satisfied.
The variation of $S_R^{(1)}$ under the gauge transformation $\delta_\Lambda^{(0)} \Phi$
is given by
\begin{equation}
\delta_\Lambda^{(0)} S_R^{(1)}
= \alpha_1 \, \langle \, Q \Lambda, \Psi^2 \, \rangle
= \alpha_1 \, \langle \, \Lambda, ( Q \Psi ) \, \Psi -\Psi \, ( Q \Psi ) \, \rangle
= {}-\alpha_1 \, \langle \, \{ \, \Psi, \Lambda \, \}, Q \Psi \, \rangle \,.
\end{equation}
This takes the form of~\eqref{Q-Psi-term}
so that this can be canceled by $\delta_\Lambda^{(1)} S_R^{(0)}$
with $\delta_\Lambda^{(1)} \Psi$ given by
\begin{equation}
\delta_\Lambda^{(1)} \Psi = {}-\alpha_1 \, X \eta \, \{ \, \Psi, \Lambda \, \} \,.
\end{equation}
The variation of $S_R^{(1)}$ under the gauge transformation $\delta_\Omega^{(0)} \Phi$
is given by
\begin{equation}
\delta_\Omega^{(0)} S_R^{(1)}
= \alpha_1 \, \langle \, \eta \Omega, \Psi^2 \, \rangle
= \alpha_1 \, \langle \, \Omega, ( \eta \Psi ) \, \Psi -\Psi \, ( \eta \Psi ) \, \rangle = 0
\end{equation}
because $\eta \Psi = 0$.
Therefore, we do not need $\delta_\Omega^{(1)} S_R^{(0)}$ and we have
\begin{equation}
\delta_\Omega^{(1)} \Psi = 0 \,.
\end{equation}
The variation of $S_R^{(1)}$ under the gauge transformation $\delta_\lambda^{(0)} \Psi$
is given by
\begin{equation}
\begin{split}
\delta_\lambda^{(0)} S_R^{(1)}
& = \alpha_1 \, \langle \, \Phi, ( Q \lambda ) \, \Psi \, \rangle
+\alpha_1 \, \langle \, \Phi, \Psi \, ( Q \lambda ) \, \rangle \\
& = {}-\alpha_1 \, \langle \, Q \Phi, \lambda \Psi \, \rangle
-\alpha_1 \, \langle \, \Phi, \lambda \, ( Q \Psi ) \, \rangle
+\alpha_1 \, \langle \, Q \Phi, \Psi \lambda \, \rangle
+\alpha_1 \, \langle \, \Phi, ( Q\Psi ) \, \lambda \, \rangle \\
& = {}-\alpha_1 \, \langle \, [ \, \Psi, \lambda \, ], Q \Phi \, \rangle
-\alpha_1 \, \langle \, [ \, \Phi, \lambda \, ], Q \Psi \, \rangle \\
& = {}-\alpha_1 \, \langle \, [ \, \Psi, \eta \Xi \lambda \, ], Q \Phi \, \rangle
-\alpha_1 \, \langle \, [ \, \Phi, \eta \Xi \lambda \, ], Q \Psi \, \rangle \\
& = \alpha_1 \, \langle \, \{ \Psi, \Xi \lambda \, \}, Q \eta \Phi \, \rangle
+\alpha_1 \, \langle \, \{ \, \eta \Phi, \Xi \lambda \, \}, Q \Psi \, \rangle \,.
\end{split}
\end{equation}
This can be canceled by $\delta_\lambda^{(1)} S_{NS}^{(0)}$ with $\delta_\lambda^{(1)} \Phi$ 
and $\delta_\lambda^{(1)} S_R^{(0)}$ with $\delta_\lambda^{(1)} \Psi$ given by
\begin{align}
\delta_\lambda^{(1)} \Phi & = \alpha_1 \, \{ \Psi, \Xi \lambda \, \} \,,
\label{delta_R^(1)-Phi-alpha_1} \\
\delta_\lambda^{(1)} \Psi & = \alpha_1 \, X \eta \, \{ \, \eta \Phi, \Xi \lambda \, \} \,.
\label{delta_R^(1)-Psi-alpha_1}
\end{align}
Note that the forms of $\delta_\lambda^{(1)} \Phi$ and $\delta_\lambda^{(1)} \Psi$ are not unique.
For example, if we instead transform $\langle \, [ \, \Psi, \lambda \, ], Q \Phi \, \rangle$ as
\begin{equation}
\langle \, [ \, \Psi, \lambda \, ], Q \Phi \, \rangle
= \langle \, \eta \xi_0 \, [ \, \Psi, \lambda \, ], Q \Phi \, \rangle
= \langle \, \xi_0 \, [ \, \Psi, \lambda \, ], Q \eta \Phi \, \rangle \,,
\end{equation}
we obtain
\begin{equation}
\tilde{\delta}_\lambda^{(1)} \Phi = {}-\alpha_1 \, \xi_0 \, [ \, \Psi, \lambda \, ] \,.
\end{equation}
However, the difference between $\delta_\lambda^{(1)} \Phi$
and $\tilde{\delta}_\lambda^{(1)} \Phi$ can be absorbed
into a correction to $\Omega$
in the gauge transformation $\delta_\Omega^{(0)} \Phi = \eta \Omega$ because
\begin{equation}
\delta_\lambda^{(1)} \Phi -\tilde{\delta}_\lambda^{(1)} \Phi
= \alpha_1 \, \{ \Psi, \Xi \lambda \, \} +\alpha_1 \, \xi_0 \, [ \, \Psi, \lambda \, ]
= \eta \, ( \, \alpha_1 \, \xi_0 \, \{ \Psi, \Xi \lambda \, \} \, ) \,.
\end{equation}
We choose the forms of $\delta_\lambda \Phi$ and $\delta_\lambda \Psi$
such that $\lambda$ appears in the combination $\Xi \lambda$ except for $Q \lambda$.
This corresponds to writing $Q \lambda$ as $Q \eta \Xi \lambda$
and transforming $\delta_\lambda^{(0)} S_R^{(1)}$ as follows:
\begin{align}
\delta_\lambda^{(0)} S_R^{(1)}
& = \alpha_1 \, \langle \, \Phi, ( Q \eta \Xi \lambda ) \, \Psi \, \rangle
+\alpha_1 \, \langle \, \Phi, \Psi \, ( Q \eta \Xi \lambda ) \, \rangle \nonumber \\
& = \alpha_1 \, \langle \, \eta \Phi, ( Q \Xi \lambda ) \, \Psi \, \rangle
-\alpha_1 \, \langle \, \eta \Phi, \Psi \, ( Q \Xi \lambda ) \, \rangle \nonumber \\
& = \alpha_1 \, \langle \, Q \eta \Phi, ( \Xi \lambda ) \Psi \, \rangle
+\alpha_1 \, \langle \, \eta \Phi, ( \Xi \lambda ) \, ( Q \Psi ) \, \rangle
+\alpha_1 \, \langle \, Q \eta \Phi, \Psi ( \Xi \lambda ) \, \rangle
-\alpha_1 \, \langle \, \eta \Phi, ( Q\Psi ) \, ( \Xi \lambda ) \, \rangle \nonumber \\
& = \alpha_1 \, \langle \, \{ \Psi, \Xi \lambda \, \}, Q \eta \Phi \, \rangle
+\alpha_1 \, \langle \, \{ \, \eta \Phi, \Xi \lambda \, \}, Q \Psi \, \rangle \,.
\end{align}
We then obtain $\delta_\lambda^{(1)} \Phi$ in~\eqref{delta_R^(1)-Phi-alpha_1}
and $\delta_\lambda^{(1)} \Psi$ in~\eqref{delta_R^(1)-Psi-alpha_1}.

\subsection{The quartic interaction}

Let us move on to the construction of the quartic interaction.
We construct $S_R^{(2)}$ such that
\begin{align}
\delta_\Lambda^{(0)} S_R^{(2)} +\delta_\Lambda^{(1)} S_R^{(1)}
+\delta_\Lambda^{(2R)} S_{NS}^{(0)} +\delta_\Lambda^{(2)} S_R^{(0)} & = 0 \,, \\
\delta_\Omega^{(0)} S_R^{(2)} +\delta_\Omega^{(1)} S_R^{(1)}
+\delta_\Omega^{(2R)} S_{NS}^{(0)} +\delta_\Omega^{(2)} S_R^{(0)} & = 0 \,, \\
\delta_\lambda^{(0)} S_R^{(2)} +\delta_\lambda^{(1)} S_{NS}^{(1)} +\delta_\lambda^{(1)} S_R^{(1)}
+\delta_\lambda^{(2)} S_{NS}^{(0)} +\delta_\lambda^{(2)} S_R^{(0)} & = 0
\end{align}
are satisfied for appropriate choices of
the parameter $\alpha_1$ appearing in $S_R^{(1)}$,
$\delta_\Lambda^{(1)} \Psi$, 
$\delta_\lambda^{(1)} \Phi$, and $\delta_\lambda^{(1)} \Psi$
and the gauge transformations
$\delta_\Lambda^{(2R)} \Phi$, $\delta_\Lambda^{(2)} \Psi$,
$\delta_\Omega^{(2R)} \Phi$, $\delta_\Omega^{(2)} \Psi$,
$\delta_\lambda^{(2)} \Phi$, and $\delta_\lambda^{(2)} \Psi$.

\subsubsection{The gauge transformation with the parameter $\Lambda$}

The variation of $S_R^{(1)}$ under the gauge transformations
$\delta_\Lambda^{(1)} \Phi$ and $\delta_\Lambda^{(1)} \Psi$ is given by
\begin{equation}
\delta_\Lambda^{(1)} S_R^{(1)}
= {}-\frac{\alpha_1}{2} \, \langle \, [ \, \Phi, Q \Lambda \, ], \Psi^2 \, \rangle
{}-\alpha_1^2 \, \langle \, \Phi, ( X \eta \, \{ \Psi, \Lambda \} ) \, \Psi \, \rangle
-\alpha_1^2 \, \langle \, \Phi, \Psi \, ( X \eta \, \{ \Psi, \Lambda \} ) \, \rangle \,.
\end{equation}
Using $X = \{ Q, \Xi \, \}$, we transform
$\langle \, \Phi, ( X \eta \, \{ \Psi, \Lambda \} ) \, \Psi \, \rangle$
as follows:
\begin{equation}
\begin{split}
\langle \, \Phi, ( X \eta \, \{ \Psi, \Lambda \} ) \, \Psi \, \rangle
& = {}-\langle \, \eta \Phi, ( \{ Q, \Xi \, \} \, \{ \Psi, \Lambda \} ) \, \Psi \, \rangle \\
& = {}-\langle \, Q \eta \Phi, ( \Xi \, \{ \Psi, \Lambda \} ) \, \Psi \, \rangle
-\langle \, \eta \Phi, ( \Xi \, \{ \Psi, \Lambda \} ) \, ( Q \Psi ) \, \rangle \\
& \quad~
-\langle \, \eta \Phi, ( \Xi \, [ \, Q \Psi, \Lambda \, ] \, ) \, \Psi \, \rangle
+\langle \, \eta \Phi, ( \Xi \, [ \, \Psi, Q \Lambda \, ] \, ) \, \Psi \, \rangle \\
& = {}-\langle \, ( \Xi \, \{ \Psi, \Lambda \} ) \, \Psi, Q \eta \Phi \, \rangle
-\langle \, ( \eta \Phi ) \, ( \Xi \, \{ \Psi, \Lambda \} ), Q \Psi \, \rangle \\
& \quad~
{}-\langle \, \{ \, \Xi \, ( \Psi ( \eta \Phi ) ), \Lambda \}, Q \Psi \, \rangle
-\langle \, Q \Lambda, \{ \, \Psi, \Xi \, ( \Psi ( \eta \Phi ) ) \, \} \, \rangle \,.
\end{split}
\end{equation}
We similarly transform
$\langle \, \Phi, \Psi \, ( X \eta \, \{ \Psi, \Lambda \} ) \, \rangle$ to find
\begin{equation}
\begin{split}
\delta_\Lambda^{(1)} S_R^{(1)}
& = \frac{\alpha_1}{2} \, \langle \, Q \Lambda, [ \, \Phi, \Psi^2 \, ] \, \rangle
+\alpha_1^2 \, \langle \, \{ \Psi, \Xi \, \{ \Psi, \Lambda \} \}, Q \eta \Phi \, \rangle
+\alpha_1^2 \, \langle \, \{ \eta \Phi, \Xi \, \{ \Psi, \Lambda \} \}, Q \Psi \, \rangle \\
& \quad~
+\alpha_1^2 \, \langle \, \{ \, \Xi \, \{ \eta \Phi, \Psi \}, \Lambda \}, Q \Psi \, \rangle
+\alpha_1^2 \, \langle \, Q \Lambda, \{ \, \Psi, \Xi \, \{ \eta \Phi, \Psi \} \} \, \rangle \,.
\label{delta_NS^(1)-S_R^(1)}
\end{split}
\end{equation}
From the structure of the last term on the right-hand side of~\eqref{delta_NS^(1)-S_R^(1)},
let us consider a quartic interaction $S_R^{(2)}$ of the form
\begin{equation}
S_R^{(2)} = \alpha_2 \, \langle \, \Phi, \{ \Psi, \Xi \, \{ \eta \Phi, \Psi \} \} \, \rangle \,,
\end{equation}
where $\alpha_2$ is a constant to be determined.
The variation of $S_R^{(2)}$ under the gauge transformation $\delta_\Lambda^{(0)} \Phi$
is given by
\begin{equation}
\begin{split}
& \delta_\Lambda^{(0)} S_R^{(2)}
= \alpha_2 \, \langle \, \delta_\Lambda^{(0)} \Phi, \{ \Psi, \Xi \, \{ \eta \Phi, \Psi \} \} \, \rangle
+\alpha_2 \, \langle \, \Phi, \{ \Psi, \Xi \, \{ \eta \, \delta_\Lambda^{(0)} \Phi, \Psi \} \} \, \rangle \\
& = \alpha_2 \, \langle \, \delta_\Lambda^{(0)} \Phi, 2 \,\{ \Psi, \Xi \, \{ \eta \Phi, \Psi \} \}
-[ \, \Phi, \Psi^2 \, ] \, \rangle \\
& = 2 \, \alpha_2 \, \langle \, Q \Lambda, \{ \Psi, \Xi \, \{ \eta \Phi, \Psi \} \} \, \rangle
-\alpha_2 \, \langle \, Q \Lambda, [ \, \Phi, \Psi^2 \, ] \, \rangle \,.
\end{split}
\end{equation}
Comparing this with~\eqref{delta_NS^(1)-S_R^(1)},
we find that the constants $\alpha_1$ and $\alpha_2$ should be chosen to be
\begin{equation}
\alpha_1 = {}-1 \,, \qquad \alpha_2 = {}-\frac{1}{2} \,,
\end{equation}
and then we have
\begin{equation}
\begin{split}
\delta_\Lambda^{(0)} S_R^{(2)} +\delta_\Lambda^{(1)} S_R^{(1)}
& = \langle \, \{ \Psi, \Xi \, \{ \Psi, \Lambda \} \}, Q \eta \Phi \, \rangle \\
& \quad~ +\langle \, \{ \eta \Phi, \Xi \, \{ \Psi, \Lambda \} \}, Q \Psi \, \rangle
+\langle \, \{ \, \Xi \, \{ \eta \Phi, \Psi \}, \Lambda \}, Q \Psi \, \rangle \,.
\end{split}
\end{equation}
The term~$\langle \, \{ \Psi, \Xi \, \{ \Psi, \Lambda \} \}, Q \eta \Phi \, \rangle$
containing $Q \eta \Phi$ can be canceled by $\delta_\Lambda^{(2R)} S_{NS}^{(0)}$
with $\delta_\Lambda^{(2R)} \Phi$ given by
\begin{equation}
\delta_\Lambda^{(2R)} \Phi = \{ \Psi, \Xi \, \{ \Psi, \Lambda \} \} \,. 
\end{equation}
The remaining terms take the form of~\eqref{Q-Psi-term}
so that they can be canceled by $\delta_\Lambda^{(2)} S_R^{(0)}$
with $\delta_\Lambda^{(2)} \Psi$ given by
\begin{equation}
\delta_\Lambda^{(2)} \Psi = X \eta \, \{ \, \Xi \, \{ \eta \Phi, \Psi \}, \Lambda \}
+ X \eta \, \{ \eta \Phi, \Xi \, \{ \Psi, \Lambda \} \} \,.
\end{equation}

\subsubsection{The gauge transformation with the parameter $\Omega$}

The variation of $S_R^{(1)}$ under the gauge transformation
$\delta_\Omega^{(1)} \Phi$ is given by
\begin{equation}
\delta_\Omega^{(1)} S_R^{(1)}
= {}-\frac{1}{2} \, \langle \, [ \, \Phi, \eta \Omega \, ], \Psi^2 \, \rangle
= \frac{1}{2} \, \langle \, \eta \Omega, [ \, \Phi, \Psi^2 \, ] \, \rangle \,.
\end{equation}
The variation of $S_R^{(2)}$ under the gauge transformation $\delta_\Omega^{(0)} \Phi$
is given by
\begin{equation}
\delta_\Omega^{(0)} S_R^{(2)}
= {}-\frac{1}{2} \, \langle \, \eta \Omega, \{ \Psi, \Xi \, \{ \eta \Phi, \Psi \} \} \, \rangle
= {}-\frac{1}{2} \, \langle \, \eta \Omega, \{ \Psi, [ \, \Phi, \Psi \, ] \, \} \, \rangle
= {}-\frac{1}{2} \, \langle \, \eta \Omega, [ \, \Phi, \Psi^2 \, ] \, \rangle \,.
\end{equation}
Since
\begin{equation}
\delta_\Omega^{(0)} S_R^{(2)} +\delta_\Omega^{(1)} S_R^{(1)} = 0 \,,
\end{equation}
we do not need
$\delta_\Omega^{(2R)} S_{NS}^{(0)}$ and $\delta_\Omega^{(2)} S_R^{(0)}$,
and we have
\begin{equation}
\delta_\Omega^{(2R)} \Phi = 0 \,, \qquad
\delta_\Omega^{(2)} \Psi = 0 \,.
\end{equation}

\subsubsection{The gauge transformation with the parameter $\lambda$}

Let us next calculate the variations $\delta_\lambda^{(1)} S_{NS}^{(1)}$,
$\delta_\lambda^{(1)} S_R^{(1)}$, and $\delta_\lambda^{(0)} S_R^{(2)}$
and express each term in the form of an inner product with $\Xi \lambda$.
The variation $\delta_\lambda^{(1)} S_{NS}^{(1)}$ is given by
\begin{equation}
\begin{split}
\delta_\lambda^{(1)} S_{NS}^{(1)}
& = {}-\frac{1}{2} \, \langle \, \delta_\lambda^{(1)} \Phi, \{ Q \Phi, \eta \Phi \} \, \rangle
= \frac{1}{2} \, \langle \, \{ \Psi, \Xi \lambda \, \}, \{ Q \Phi, \eta \Phi \} \, \rangle \\
& = {}-\frac{1}{2} \, \langle \, \Xi \lambda, [ \, \{ Q \Phi, \eta \Phi \}, \Psi \, ] \, \rangle \,.
\end{split}
\end{equation}
The variation $\delta_\lambda^{(1)} S_R^{(1)}$ is given by
\begin{equation}
\begin{split}
\delta_\lambda^{(1)} S_R^{(1)}
= \langle \, \{ \Psi, \Xi \lambda \, \}, \Psi^2 \, \rangle
+\langle \, \Phi, ( X \eta \, \{ \, \eta \Phi, \Xi \lambda \, \} \, ) \, \Psi \, \rangle
+\langle \, \Phi, \Psi \, ( X \eta \, \{ \, \eta \Phi, \Xi \lambda \, \} \, ) \, \rangle \,.
\end{split}
\end{equation}
The first term on the right-hand side vanishes:
\begin{equation}
\langle \, \{ \Psi, \Xi \lambda \, \}, \Psi^2 \, \rangle
= {}-\langle \, \Xi \lambda, \Psi^3 \, \rangle
+\langle \, \Xi \lambda, \Psi^3 \, \rangle = 0 \,.
\end{equation}
The remaining terms are
\begin{equation}
\begin{split}
\delta_\lambda^{(1)} S_R^{(1)}
& = {}-\langle \, \eta \Phi, ( \{ Q, \Xi \} \, \{ \, \eta \Phi, \Xi \lambda \, \} \, ) \, \Psi \, \rangle
+\langle \, \eta \Phi, \Psi \, ( \{ Q, \Xi \} \, \{ \, \eta \Phi, \Xi \lambda \, \} \, ) \, \rangle \\
& = \langle \, \Xi \lambda, [ \, \eta \Phi, \{ Q, \Xi \} \, \{ \, \eta \Phi, \Psi \} \, ] \, \rangle \,.
\end{split}
\end{equation}
For the variation $\delta_\lambda^{(0)} S_R^{(2)}$,
we write $Q \lambda$ as $Q \eta \Xi \lambda$ and find
\begin{equation}
\begin{split}
\delta_\lambda^{(0)} S_R^{(2)}
& = {}-\frac{1}{2} \, \langle \, \Phi, \{ Q \eta \Xi \lambda, \Xi \, \{ \eta \Phi, \Psi \} \} \, \rangle
-\frac{1}{2} \, \langle \, \Phi, \{ \Psi, \Xi \, \{ \eta \Phi, Q \eta \Xi \lambda \} \} \, \rangle \\
& = \frac{1}{2} \, \langle \, Q \eta \Xi \lambda,
[ \, \Phi, \Xi \, \{ \eta \Phi, \Psi \} \, ] +[ \, \Xi \, [ \, \Psi, \Phi \, ], \eta \Phi \, ] \, \rangle \\
& = \langle \, Q \Xi \lambda, \{ \eta \Phi, \Xi \, \{ \eta \Phi, \Psi \} \} \, \rangle
+\frac{1}{2} \, \langle \, Q \Xi \lambda, \{ \, [ \, \Phi, \eta \Phi \, ], \Psi \} \, \rangle \\
& = \langle \, \Xi \lambda, Q \, \{ \eta \Phi, \Xi \, \{ \eta \Phi, \Psi \} \} \, \rangle
+\frac{1}{2} \, \langle \, \Xi \lambda, Q \, \{ \, [ \, \Phi, \eta \Phi \, ], \Psi \} \, \rangle \,,
\end{split}
\end{equation}
where in an intermediate step we used the following Jacobi identity:
\begin{equation}
[ \, \Phi, \{ \eta \Phi, \Psi \} \, ] +\{ \, [ \, \Psi, \Phi \, ], \eta \Phi \}
= \{ \, [ \, \Phi, \eta \Phi \, ], \Psi \} \,.
\end{equation}
We then find
\begin{equation}
\begin{split}
& \delta_\lambda^{(1)} S_{NS}^{(1)} +\delta_\lambda^{(1)} S_R^{(1)} +\delta_\lambda^{(0)} S_R^{(2)} \\
& = {}{}-\frac{1}{2} \, \langle \, \Xi \lambda, [ \, \{ Q \Phi, \eta \Phi \}, \Psi \, ] \, \rangle
+\langle \, \Xi \lambda, [ \, \eta \Phi, \{ Q, \Xi \} \, \{ \, \eta \Phi, \Psi \} \, ] \, \rangle \\
& \quad~
+\langle \, \Xi \lambda, Q \, \{ \eta \Phi, \Xi \, \{ \eta \Phi, \Psi \} \} \, \rangle
+\frac{1}{2} \, \langle \, \Xi \lambda, Q \, \{ \, [ \, \Phi, \eta \Phi \, ], \Psi \} \, \rangle \\
& = \langle \, \Xi \lambda, [ \, Q \eta \Phi, \Xi \, \{ \eta \Phi, \Psi \} \, ] \, \rangle
+\langle \, \Xi \lambda, [ \, \eta \Phi, \Xi \, [ \, Q \eta \Phi, \Psi \, ] \, ] \, \rangle
-\langle \, \Xi \lambda, [ \, \eta \Phi, \Xi \, [ \, \eta \Phi, Q \Psi \, ] \, ] \, \rangle \\
& \quad~
+\frac{1}{2} \, \langle \, \Xi \lambda, [ \, [ \, \Phi, Q \eta \Phi \, ], \Psi \, ] \, \rangle
-\frac{1}{2} \, \langle \, \Xi \lambda, [ \, [ \, \Phi, \eta \Phi \, ], Q \Psi \, ] \, \rangle \\
& = {}-\langle \, \{ \Psi, \Xi \, \{ \eta \Phi, \Xi \lambda \} \}
+\{ \, \Xi \, \{ \eta \Phi, \Psi \}, \Xi \lambda \}, Q \eta \Phi \, \rangle
+\frac{1}{2} \, \langle \, [ \, \Phi, \{ \Psi, \Xi \lambda \} \, ], Q \eta \Phi \, \rangle \\
& \quad~
{}-\langle \, \{ \, \eta \Phi, \Xi \, \{ \eta \Phi, \Xi \lambda \} \}, Q \Psi \, \rangle
-\frac{1}{2} \, \langle \, \{ [ \, \Phi, \eta \Phi \, ], \Xi \lambda \}, Q \Psi \, \rangle \,.
\end{split}
\end{equation}
These terms are canceled by $\delta_\lambda^{(2)} S_{NS}^{(0)}$
and $\delta_\lambda^{(2)} S_R^{(0)}$
with $\delta_\lambda^{(2)} \Phi$ and $\delta_\lambda^{(2)} \Psi$
given by
\begin{align}
\delta_\lambda^{(2)} \Phi
& = {}-\{ \Psi, \Xi \, \{ \eta \Phi, \Xi \lambda \} \}
-\{ \, \Xi \, \{ \eta \Phi, \Psi \}, \Xi \lambda \}
+\frac{1}{2} \, [ \, \Phi, \{ \Psi, \Xi \lambda \} \, ] \,, \\
\delta_\lambda^{(2)} \Psi
& = {}-X \eta \, \{ \, \eta \Phi, \Xi \, \{ \eta \Phi, \Xi \lambda \} \}
-\frac{1}{2} \, X \eta \, \{ [ \, \Phi, \eta \Phi \, ], \Xi \lambda \} \,.
\end{align}

\subsection{Summary}\label{summary}

Let us summarize the results of this section.
The action in the NS sector is given by
\begin{equation}
S_{NS} = S_{NS}^{(0)} +g \, S_{NS}^{(1)} +g^2 \, S_{NS}^{(2)} +O(g^3) \,,
\end{equation}
where
\begin{align}
S_{NS}^{(0)} & = {}-\frac{1}{2} \, \langle \, \Phi, Q \eta \Phi \, \rangle \,, \\
S_{NS}^{(1)} & = {}-\frac{1}{6} \, \langle \, \Phi, Q \, [ \, \Phi, \eta \Phi \, ] \, \rangle \,, \\
S_{NS}^{(2)} &
= {}-\frac{1}{24} \, \langle \, \Phi, Q \, [ \, \Phi, [ \, \Phi, \eta \Phi \, ] \, ] \, \rangle \,.
\end{align}
The action in the Ramond sector is given by
\begin{equation}
S_R = S_R^{(0)} +g \, S_R^{(1)} +g^2 \, S_R^{(2)} +O(g^3) \,,
\end{equation}
where
\begin{align}
S_R^{(0)} & = {}-\frac{1}{2} \, \llangle \, \Psi, Y Q \Psi \, \rrangle \,, \\
S_R^{(1)} & = {}-\langle \, \Phi, \Psi^2 \, \rangle \,, \\
S_R^{(2)} & = {}-\frac{1}{2} \, \langle \, \Phi, \{ \Psi, \Xi \, \{ \eta \Phi, \Psi \} \} \, \rangle \,.
\end{align}
The gauge transformation
with the gauge parameter $\Lambda$ in the NS sector
is given by
\begin{align}
\delta_\Lambda \Phi
& = \delta_\Lambda^{(0)} \Phi +g \, \delta_\Lambda^{(1)} \Phi +g^2 \, \delta_\Lambda^{(2)} \Phi +O(g^3) \,, \\
\delta_\Lambda \Psi
& = \delta_\Lambda^{(0)} \Psi +g \, \delta_\Lambda^{(1)} \Psi +g^2 \, \delta_\Lambda^{(2)} \Psi +O(g^3) \,,
\end{align}
where
\begin{align}
\delta_\Lambda^{(0)} \Phi & = Q \Lambda \,, \\
\delta_\Lambda^{(1)} \Phi
& = {}-\frac{1}{2} \, [ \, \Phi, Q \Lambda \, ] \,, \\
\delta_\Lambda^{(2)} \Phi
& = \frac{1}{12} \, [ \, \Phi, [ \, \Phi, Q \Lambda \, ] \, ]
+\{ \Psi, \Xi \, \{ \Psi, \Lambda \} \} \,, \\
\delta_\Lambda^{(0)} \Psi & = 0 \,, \\
\delta_\Lambda^{(1)} \Psi & = \, X \eta \, \{ \, \Psi, \Lambda \, \} \,, \\
\delta_\Lambda^{(2)} \Psi & = X \eta \, \{ \, \Xi \, \{ \eta \Phi, \Psi \}, \Lambda \}
+ X \eta \, \{ \eta \Phi, \Xi \, \{ \Psi, \Lambda \} \} \,.
\end{align}
The gauge transformation
with the gauge parameter $\Omega$ in the NS sector is given by
\begin{align}
\delta_\Omega \Phi
& = \delta_\Omega^{(0)} \Phi +g \, \delta_\Omega^{(1)} \Phi +g^2 \, \delta_\Omega^{(2)} \Phi +O(g^3) \,, \\
\delta_\Omega \Psi
& = \delta_\Omega^{(0)} \Psi +g \, \delta_\Omega^{(1)} \Psi +g^2 \, \delta_\Omega^{(2)} \Psi +O(g^3) \,,
\end{align}
where
\begin{align}
\delta_\Omega^{(0)} \Phi & = \eta \Omega \,, \\
\delta_\Omega^{(1)} \Phi
& = \frac{1}{2} \, [ \, \Phi, \eta \Omega \, ] \,, \\
\delta_\Omega^{(2)} \Phi
& = \frac{1}{12} \, [ \, \Phi, [ \, \Phi, \eta \Omega \, ] \, ] \,, \\
\delta_\Omega^{(0)} \Psi & = 0 \,, \\
\delta_\Omega^{(1)} \Psi & = 0 \,, \\
\delta_\Omega^{(2)} \Psi & = 0 \,.
\end{align}
The gauge transformation
with the gauge parameter $\lambda$ in the Ramond sector is given by
\begin{align}
\delta_\lambda \Phi
& = \delta_\lambda^{(0)} \Phi +g \, \delta_\lambda^{(1)} \Phi +g^2 \, \delta_\lambda^{(2)} \Phi +O(g^3) \,, \\
\delta_\lambda \Psi
& = \delta_\lambda^{(0)} \Psi +g \, \delta_\lambda^{(1)} \Psi +g^2 \, \delta_\lambda^{(2)} \Psi +O(g^3) \,,
\end{align}
where
\begin{align}
\delta_\lambda^{(0)} \Phi & = 0 \,, \\
\delta_\lambda^{(1)} \Phi & = {}-\{ \Psi, \Xi \lambda \, \} \,, \\
\delta_\lambda^{(2)} \Phi
& = {}-\{ \Psi, \Xi \, \{ \eta \Phi, \Xi \lambda \} \}
-\{ \, \Xi \, \{ \eta \Phi, \Psi \}, \Xi \lambda \}
+\frac{1}{2} \, [ \, \Phi, \{ \Psi, \Xi \lambda \} \, ] \,, \\
\delta_\lambda^{(0)} \Psi & = Q \lambda \,, \\
\delta_\lambda^{(1)} \Psi & = {}-X \eta \, \{ \, \eta \Phi, \Xi \lambda \, \} \,, \\
\delta_\lambda^{(2)} \Psi
& = {}-X \eta \, \{ \, \eta \Phi, \Xi \, \{ \eta \Phi, \Xi \lambda \} \}
-\frac{1}{2} \, X \eta \, \{ [ \, \Phi, \eta \Phi \, ], \Xi \lambda \} \,.
\end{align}

\section{Complete action}\label{complete-action-section}
\setcounter{equation}{0}

In this section we present a complete action.
We derive the equations of motion
and show the gauge invariance of the action.

\subsection{Action and gauge transformations}

The complete action $S$ is given by
\begin{equation}
 S = {}-\frac{1}{2} \, \llangle \, \Psi, YQ\Psi \, \rrangle
-\int_0^1 dt \, \langle \, A_t(t), QA_\eta(t)+ ( \, F(t) \Psi \, )^2 \, \rangle \,,
\label{complete-action}
\end{equation}
where 
\begin{equation}
\begin{split}
F(t) \Psi  & = \Psi + \Xi \, \{A_\eta(t),\Psi \} 
+\Xi \, \{A_\eta(t), \Xi \, \{ A_\eta(t),\Psi\}\} +\cdots \\
& = \sum_{n=0}^\infty \, \underbrace{\Xi \, \{A_\eta(t), \Xi \, \{A_\eta(t), \cdots, \Xi \, \{
A_\eta(t)}_{n},\Psi \} \cdots\}\} \,,
\end{split}
\end{equation}
and the string fields $A_\eta(t)$ and $A_t(t)$ satisfy the relations
\begin{equation}
\eta A_\eta (t) = A_\eta (t) \, A_\eta (t) \,, \qquad
\partial_t A_\eta (t) = \eta A_t (t) -A_\eta (t) \, A_t (t) +A_t (t) \, A_\eta (t)
\label{fundamental-relations}
\end{equation}
with $A_\eta (0) = 0$ and $A_t (0) = 0$.
We can parameterize $A_\eta (t)$ and $A_t (t)$ satisfying~\eqref{fundamental-relations}
in terms of $\Phi (t)$ in the NS sector with $\Phi (0) = 0$ as
\begin{equation}
A_\eta (t) = ( \, \eta e^{\Phi(t)} \, ) \, e^{-\Phi(t)} \,, \qquad
A_t (t) = ( \, \partial_t e^{\Phi(t)} \, ) \, e^{-\Phi(t)} \,.
\label{Berkovits-parameterization}
\end{equation}
The string field $\Phi (t)$ is a Grassmann-even state
and is in the large Hilbert space.
Its ghost number is $0$ and its picture number is also $0$.
The string field $\Psi$ is in the Ramond sector.
It is a Grassmann-odd state,
its ghost number is $1$, and its picture number is $-1/2$.
It is in the small Hilbert space and is in the restricted space:
\begin{equation}
\eta \Psi = 0 \,, \qquad  X Y \Psi = \Psi \,.
\end{equation}
Note that $\Psi$ is not a function of $t$.
As we will show, the dependence of the action on $t$ is topological,
and the action is a functional of $\Phi$ and $\Psi$,
where $\Phi$ is the value of $\Phi (t)$ at $t=1$.

We will show that the action~\eqref{complete-action} is invariant
under the following gauge transformations:
\begin{subequations}
\label{gauge-transformations}
\begin{align}
A_\delta & = Q \Lambda +D_\eta \Omega
+ \{ F \Psi, F \Xi\ ( \, \{ F \Psi, \Lambda \} -\lambda \, ) \} \,, \\
\delta \Psi & = Q\lambda
+X\eta \, F \Xi \, D_\eta \, ( \, \{ F \Psi, \Lambda \} -\lambda \, ) \,,
\label{Ramond-gauge-transformation}
\end{align}
\end{subequations}
where $\Lambda$ and $\Omega$ are gauge parameters in the NS sector
and $\lambda$ is a gauge parameter in the Ramond sector satisfying
\begin{equation}
\eta \lambda = 0 \,, \qquad X Y \lambda = \lambda \,.
\end{equation}
The action of $D_\eta$ is defined by
\begin{equation}
D_\eta A = \eta A -A_\eta \, A+(-1)^A \, A \, A_\eta \,,
\label{D_eta}
\end{equation}
where
\begin{equation}
A_\eta = A_\eta (1) \,,
\end{equation}
and the action of $F$ is defined by
\begin{equation}
\begin{split}
F A & = A +\Xi \, [ \, A_\eta, A \, ]
+\Xi \, [ \, A_\eta, \Xi \, [ \, A_\eta, A \, ] \, ] + \cdots \\
& = \sum_{n=0}^\infty \,
\underbrace{\Xi \, [ \, A_\eta, \Xi \, [ \, A_\eta, \cdots \,, \Xi \, [ \, A_\eta \,}_n, A \, ] \cdots \, ] \, ]
\end{split}
\label{F-even-definition}
\end{equation}
when $A$ is a Grassmann-even state and
\begin{equation}
\begin{split}
F A & = A +\Xi \, \{ \, A_\eta, A \, \}
+\Xi \, \{ \, A_\eta, \Xi \, \{ \, A_\eta, A \, \} \, \} + \cdots \\
& = \sum_{n=0}^\infty \,
\underbrace{\Xi \, \{ \, A_\eta, \Xi \, \{ \, A_\eta, \cdots \,, \Xi \, \{ \, A_\eta \,}_n, A \, \} \cdots \, \} \, \}
\end{split}
\label{F-odd-definition}
\end{equation}
when $A$ is a Grassmann-odd state.
The string field $A_\delta$ is related to $A_\eta$ as
\begin{equation}
\delta A_\eta = D_\eta A_\delta = \eta A_\delta -[ \, A_\eta, A_\delta \, ] \,.
\end{equation}
This relation defines $A_\delta$ up to terms that are annihilated by $D_\eta$,
and the ambiguity can be absorbed by the gauge parameter $\Omega$.
For the parameterization of $A_\eta (t)$ in~\eqref{Berkovits-parameterization},
an explicit form of $A_\delta$ is
\begin{equation}
A_\delta = ( \, \delta e^\Phi \, ) \, e^{-\Phi} \,.
\end{equation}
Note that $\delta \Psi$ in~\eqref{Ramond-gauge-transformation}
is in the small Hilbert space and in the restricted space:
\begin{equation}
\eta \, \delta \Psi = 0 \,, \qquad X Y \delta \Psi = \delta \Psi \,.
\end{equation}

When we set $\Psi = 0$, the action~\eqref{complete-action}
coincides with the WZW-like action $S_{\rm WZW}$
of the Berkovits formulation~\cite{Berkovits:1995ab}:
\begin{equation}
S_{\rm WZW} = \frac{1}{2} \, \langle \, e^{-\Phi}Qe^\Phi, e^{-\Phi} \eta e^\Phi \, \rangle
-\frac{1}{2} \int^1_0 dt \, \langle \, e^{-\Phi (t)} \partial_t e^{\Phi (t)}, \,
\{ \, e^{-\Phi (t)} Q e^{\Phi (t)}, e^{-\Phi (t)} \eta e^{\Phi (t)} \, \} \, \rangle \,,
\label{S_WZW-1}
\end{equation}
and the form of $S_{\rm WZW}$ given by
\begin{equation}
S_{\rm WZW} = {}-\int_0^1 dt \, \langle \, A_t(t), QA_\eta(t) \, \rangle
\label{S_WZW-2}
\end{equation}
was recently used in~\cite{Erler:2015rra}.
While the NS sector of the action is based on the large Hilbert space,
we can apply the partial gauge fixing discussed in~\cite{Iimori:2013kha}
and obtain a gauge-invariant action based on the small Hilbert space
both for the NS sector and the Ramond sector.

The action up to quartic interactions
in Section~\ref{cubic-quartic-section}
with $g=1$ coincides with~\eqref{complete-action}
under the parameterization~\eqref{Berkovits-parameterization}.
However, the gauge invariance of the action does not depend
on this particular parameterization,
and other parameterizations of $A_\eta (t)$ and $A_t (t)$
satisfying~\eqref{fundamental-relations} are possible.
For example, as was demonstrated in~\cite{Erler:2015rra},
we can parameterize $A_\eta (t)$ and $A_t (t)$
in terms of a string field in the small Hilbert space
so that the action in the NS sector coincides
with the action constructed in~\cite{Erler:2013xta} with the $A_\infty$ structure.
Therefore, we can also regard the action~\eqref{complete-action}
as the inclusion of the Ramond sector to the action in~\cite{Erler:2013xta}.

\subsection{Algebraic ingredients}

In the rest of this section,
we derive the equations of motion from the action~\eqref{complete-action}
and show its gauge invariance.
The starting point of our discussion is the relation
\begin{equation}
\eta A_\eta (t) = A_\eta (t) \, A_\eta (t) \,.
\label{A_eta-relation}
\end{equation}
This is analogous to the equation of motion $Q A +A^2 = 0$ in open bosonic string field theory,
and the string field $A_\eta (t)$ satisfying this relation corresponds to a pure gauge
with respect to the gauge transformation generated by $\eta$.
We define the covariant derivative $D_\eta (t)$ by
\begin{equation}
D_\eta (t) A = \eta A -A_\eta (t) A +(-1)^A \, A \, A_\eta (t) \,.
\end{equation}
This is a generalization of $D_\eta$ in~\eqref{D_eta},
and $D_\eta$ corresponds to $D_\eta (t)$ with $t=1$.
The covariant derivative $D_\eta (t)$ squares to zero,
\begin{equation}
D_\eta (t)^2 = 0 \,,
\end{equation}
because of the relation~\eqref{A_eta-relation}.
It acts as a derivation with respect to the star product,
\begin{equation}
D_\eta (t) ( A \, B \, )
= ( D_\eta (t) \, A \, ) \, B
+(-1)^A \, A \, ( D_\eta (t) \, B \, ) \,,
\end{equation}
and it is BPZ odd:
\begin{equation}
\langle \, D_\eta (t) \, A, B \, \rangle
= {}-(-1)^A \, \langle \, A, D_\eta (t) \, B \, \rangle
\end{equation}
for any states $A$ and $B$.
The covariant derivative $D_\eta (t)$ is an important ingredient in our construction.

Another important ingredient is the linear map $F (t)$.
It is a generalization of $F$ defined in~\eqref{F-even-definition} and~\eqref{F-odd-definition},
and the action of $F (t)$ on a state $A$ in the Ramond sector is defined by
\begin{equation}
\begin{split}
F(t) A & = A +\Xi \, [ \, A_\eta(t), A \, ]
+\Xi \, [ \, A_\eta(t), \Xi \, [ \, A_\eta(t), A \, ] \, ] + \cdots \\
& = \sum_{n=0}^\infty \,
\underbrace{\Xi \, [ \, A_\eta(t), \Xi \, [ \, A_\eta(t), \cdots \,, \Xi \, [ \, A_\eta(t)}_n, A \, ] \cdots \, ] \, ]
\end{split}
\end{equation}
when $A$ is a Grassmann-even state and
\begin{equation}
\begin{split}
F(t) A & = A +\Xi \, \{ \, A_\eta(t), A \, \}
+\Xi \, \{ \, A_\eta(t), \Xi \, \{ \, A_\eta(t), A \, \} \, \} + \cdots \\
& = \sum_{n=0}^\infty \,
\underbrace{\Xi \, \{ \, A_\eta(t), \Xi \, \{ \, A_\eta(t), \cdots \,, \Xi \, \{ \, A_\eta(t)}_n, A \, \} \cdots \, \} \, \}
\end{split}
\end{equation}
when $A$ is a Grassmann-odd state.
The map $F$ in~\eqref{F-even-definition} and~\eqref{F-odd-definition}
corresponds to $F(t)$ with $t=1$.
It is useful to consider the inverse map $F^{-1} (t)$ given by
\begin{equation}
F^{-1} (t) A = A -\Xi \, ( \, A_\eta (t) \, A -(-1)^A A \, A_\eta (t) \, ) \,.
\end{equation}
Since
\begin{equation}
\begin{split}
A -\Xi \, ( \, A_\eta (t) \, A -(-1)^A A \, A_\eta (t) \, )
= A +\Xi D_\eta (t) A -\Xi \eta A
=  \eta \Xi A +\Xi D_\eta (t) A \,,
\end{split}
\end{equation}
we find
\begin{equation}
F^{-1} (t) = \eta \Xi +\Xi D_\eta (t) \,.
\label{F^-1-expression}
\end{equation}
It follows from $\eta^2 = 0$ and $D_\eta (t)^2 = 0$ that
\begin{equation}
\eta F^{-1} (t) = \eta \Xi D_\eta (t) \,, \qquad
F^{-1} (t) D_\eta (t) = \eta \Xi D_\eta (t) \,.
\end{equation}
We thus obtain
\begin{equation}
\eta F^{-1} (t) = F^{-1} (t) D_\eta (t) \,.
\end{equation}
In terms of $F (t)$, we have
\begin{equation}
D_\eta (t) F (t) = F (t) \, \eta \,.
\label{D_eta-F}
\end{equation}
An important relation can be obtained
when we multiply both sides of~\eqref{F^-1-expression} by $F (t)$:
\begin{equation}
1 = F (t) \eta \Xi +F (t) \Xi  D_\eta (t) = D_\eta (t) F (t) \Xi +F (t) \Xi  D_\eta (t) \,.
\end{equation}
We thus find
\begin{equation}
\{ \, D_\eta (t), F (t) \, \Xi \, \} = 1 \,.
\label{D_eta-F-Xi}
\end{equation}
Therefore, any state $A$ in the Ramond sector annihilated by $D_\eta (t)$,
\begin{equation}
D_\eta (t) A = 0 \,,
\end{equation}
can be written as
\begin{equation}
A = \{ \, D_\eta (t), F (t) \, \Xi \, \} A = D_\eta (t) F (t) \, \Xi \, A \,. 
\end{equation}

While we use $F (t)$ in the construction of the action in the Ramond sector,
it will be convenient to introduce $f (t)$, which acts on a state in the NS sector and satisfies
\begin{equation}
\{ \, D_\eta (t), f (t) \, \xi_0 \, \} = 1 \,.
\end{equation}
The action of $f (t)$ on a state $A$ in the NS sector is defined by
\begin{equation}
\begin{split}
f(t) A & = A +\xi_0 \, [ \, A_\eta(t), A \, ]
+\xi_0 \, [ \, A_\eta(t), \xi_0 \, [ \, A_\eta(t), A \, ] \, ] + \cdots \\
& = \sum_{n=0}^\infty \,
\underbrace{\xi_0 \, [ \, A_\eta(t), \xi_0 \, [ \, A_\eta(t), \cdots \,, \xi_0 \, [ \, A_\eta(t)}_n, A \, ] \cdots \, ] \, ]
\end{split}
\end{equation}
when $A$ is a Grassmann-even state and
\begin{equation}
\begin{split}
f(t) A & = A +\xi_0 \, \{ \, A_\eta(t), A \, \}
+\xi_0 \, \{ \, A_\eta(t), \xi_0 \, \{ \, A_\eta(t), A \, \} \, \} + \cdots \\
& = \sum_{n=0}^\infty \,
\underbrace{\xi_0 \, \{ \, A_\eta(t), \xi_0 \, \{ \, A_\eta(t), \cdots \,, \xi_0 \, \{ \, A_\eta(t)}_n, A \, \} \cdots \, \} \, \}
\end{split}
\end{equation}
when $A$ is a Grassmann-odd state.

The string fields $Q A_\eta (t)$ and $F (t) \Psi$ in the action~\eqref{complete-action}
are annihilated by $D_\eta (t)$:
\begin{align}
D_\eta (t) \, Q A_\eta (t) & = 0 \,,
\label{Q-A_eta-closed} \\
D_\eta (t) F (t) \Psi & = 0 \,.
\label{F-Psi-closed}
\end{align}
The first relation~\eqref{Q-A_eta-closed} follows from~\eqref{A_eta-relation},
and the second relation~\eqref{F-Psi-closed} follows from~\eqref{D_eta-F} and $\eta \Psi = 0$:
\begin{equation}
D_\eta (t) F (t) \Psi = F (t) \, \eta \Psi = 0 \,.
\end{equation}
The string field $\partial_t A_\eta (t)$ is also annihilated by $D_\eta (t)$:
\begin{equation}
D_\eta (t) \, \partial_t A_\eta (t) = 0 \,,
\end{equation}
which again follows from the relation~\eqref{A_eta-relation}.
Therefore, $\partial_t A_\eta (t)$ can be written as
\begin{equation}
\partial_t A_\eta (t) = D_\eta (t) A_t (t) \,,
\label{A_t-relation}
\end{equation}
where $A_t (t)$ is a string field of ghost number $0$ and picture number $0$.
Since $A_\eta (t)$ is a pure gauge for any $t$,
an infinitesimal change in $t$ should be implemented
by a gauge transformation, and $A_t (t)$ corresponds to the gauge parameter.
One choice of $A_t (t)$ is $f (t) \, \xi_0 \, \partial_t A_\eta (t)$,
but it is not unique.
Suppose that $A_t^{(1)} (t)$ and $A_t^{(2)} (t)$ both satisfy~\eqref{A_t-relation}:
\begin{equation}
\partial_t A_\eta (t) = D_\eta (t) A_t^{(1)} (t) \,, \qquad
\partial_t A_\eta (t) = D_\eta (t) A_t^{(2)} (t) \,.
\end{equation}
Then the difference $\Delta A_t (t) = A_t^{(1)} (t) -A_t^{(2)} (t)$ is annihilated by $D_\eta (t)$:
\begin{equation}
D_\eta (t) \, \Delta A_t (t)
= D_\eta (t) \, ( \, A_t^{(1)} (t) -A_t^{(2)} (t) \, ) = 0 \,.
\end{equation}
The string fields $A_\eta (t)$ and $A_t (t)$ in the action have to satisfy~\eqref{A_t-relation}.
The ambiguity in $A_t (t)$, however, does not affect the action because
\begin{equation}
\begin{split}
& \langle \, \Delta A_t (t), Q A_\eta (t) +( F (t) \Psi )^2 \, \rangle
= \langle \, \{ D_\eta (t), f (t) \xi_0 \} \, \Delta A_t (t), Q A_\eta (t) +( F (t) \Psi )^2 \, \rangle \\
& = \langle \, f (t) \xi_0 \, D_\eta (t) \, \Delta A_t (t), Q A_\eta (t) +( F (t) \Psi )^2 \, \rangle
+\langle \, f (t) \xi_0 \, \Delta A_t (t), D_\eta (t) \, Q A_\eta (t) \, \rangle \\
& \quad~
+\langle \, f (t) \xi_0 \, \Delta A_t (t), 
( D_\eta (t) F (t) \Psi ) ( F (t) \Psi ) -( F (t) \Psi ) ( D_\eta (t) F (t) \Psi ) \, \rangle
= 0 \,.
\end{split}
\end{equation}

When we parameterize $A_\eta (t)$ in terms of $\Phi (t)$,
the variation $\delta A_\eta (t)$ under $\delta \Phi (t)$
is annihilated by $D_\eta (t)$:
\begin{equation}
D_\eta (t) \, \delta A_\eta (t) = 0 \,.
\end{equation}
This follows from~\eqref{A_eta-relation},
and the underlying reason is the same as in the case of $\partial_t A_\eta (t)$.
The string field $A_\eta (t)$ is a pure gauge for any $\Phi (t)$,
and an infinitesimal change in $\Phi (t)$ should be implemented
by a gauge transformation.
We write
\begin{equation}
\delta A_\eta (t) = D_\eta (t) \, A_\delta (t) \,,
\label{A_delta-relation}
\end{equation}
where $A_\delta (t)$ corresponds to the gauge parameter.
When $A_\eta (t)$ is given,
the gauge parameter $A_\delta (t)$ satisfying~\eqref{A_delta-relation}
is again not unique,
but we only use the relation~\eqref{A_delta-relation}.

We will also need a relation between $\delta A_t (t)$ and $\partial_t A_\delta (t)$.
First, consider $\delta \, \partial_t A_\eta (t)$ using~\eqref{A_t-relation}.
We find
\begin{equation}
\delta \, \partial_t A_\eta (t)
= \delta D_\eta (t) \, A_t (t)
= [ \, \delta, D_\eta (t) \, ] \, A_t (t) +D_\eta (t) \, \delta A_t (t) \,,
\end{equation}
where the action of $[ \, \delta, D_\eta (t) \, ]$ is defined by
\begin{equation}
[ \, \delta, D_\eta (t) \, ] \, A= \delta D_\eta (t) A -D_\eta (t) \, \delta A \,,
\end{equation}
and we have
\begin{equation}
\begin{split}
[ \, \delta, D_\eta (t) \, ] \, A
& = {}-( \delta A_\eta (t) ) \, A +(-1)^A A \, ( \delta A_\eta (t) ) \\
& = {}-( D_\eta (t) A_\delta (t) ) \, A +(-1)^A A \, ( D_\eta (t) A_\delta (t) ) \,.
\label{delta-D_eta}
\end{split}
\end{equation}
Therefore, $\delta \, \partial_t A_\eta (t)$ is given by
\begin{equation}
\delta \, \partial_t A_\eta (t)
= D_\eta (t) \, \delta A_t (t) -[ \, D_\eta (t) \, A_\delta (t), A_t (t) \, ] \,.
\end{equation}
Second, consider $\partial_t \, \delta A_\eta (t)$ using~\eqref{A_delta-relation}.
We find
\begin{equation}
\partial_t \, \delta A_\eta (t)
= \partial_t \, D_\eta (t) \, A_\delta (t)
= [ \, \partial_t, D_\eta (t) \, ] \, A_\delta (t) +D_\eta (t) \, \partial_t A_\delta (t) \,,
\end{equation}
where the action of $[ \, \partial_t, D_\eta (t) \, ]$ is defined by
\begin{equation}
[ \, \partial_t, D_\eta (t) \, ] \, A = \partial_t D_\eta (t) A -D_\eta (t) \, \partial_t A \,,
\end{equation}
and we have
\begin{equation}
\begin{split}
[ \, \partial_t, D_\eta (t) \, ] \, A
& = {}-( \partial_t A_\eta (t) ) \, A +(-1)^A \, A \, ( \partial_t A_\eta (t) ) \\
& = {}-( D_\eta (t) A_t (t) ) \, A +(-1)^A \, A \, ( D_\eta (t) A_t (t) ) \,.
\label{partial_t-D_eta}
\end{split}
\end{equation}
Therefore, $\partial_t \, \delta A_\eta (t)$ is given by
\begin{equation}
\partial_t \, \delta A_\eta (t)
= D_\eta (t) \, \partial_t A_\delta (t) -[ \, D_\eta (t) \, A_t (t), A_\delta (t) \, ] \,.
\end{equation}
Since $\delta \, \partial_t A_\eta (t) -\partial_t \, \delta A_\eta (t) = 0$, we find
\begin{equation}
\begin{split}
& D_\eta (t) \, \delta A_t (t) -[ \, D_\eta (t) \, A_\delta (t), A_t (t) \, ]
-D_\eta (t) \, \partial_t A_\delta (t) -[ \, A_\delta (t), D_\eta (t) \, A_t (t) \, ] \\
& = D_\eta (t) \, ( \, \delta A_t (t) -\partial_t A_\delta (t) -[ \, \, A_\delta (t), A_t (t) \, ] \, ) = 0 \,.
\end{split}
\end{equation}
We write this as
\begin{equation}
D_\eta (t) F_{\delta t} (t) = 0 \,,
\end{equation}
where
\begin{equation}
F_{\delta t} (t) = \delta A_t (t) -\partial_t A_\delta (t)
-[ \, A_\delta (t), A_t (t) \, ] \,.
\label{F_delta-t}
\end{equation}
When we parameterize $A_\eta (t)$ as $A_\eta (t) = ( \, \eta e^{\Phi (t)} \, ) \, e^{-\Phi (t)}$
and choose $A_t (t)$ and $A_\delta (t)$ to be
\begin{equation}
A_t (t) = ( \, \partial_t e^{\Phi (t)} \, ) \, e^{-\Phi (t)} \,, \qquad
A_\delta (t) = ( \, \delta e^{\Phi (t)} \, ) \, e^{-\Phi (t)} \,,
\end{equation}
the string field $F_{\delta t} (t)$ vanishes.
In general, however, this is not the case,
and in fact it was found in~\cite{Erler:2015rra} that
$F_{\delta t} (t)$ is nonvanishing
for the parameterization of $A_\eta (t)$ and $A_t (t)$
to reproduce the action with the $A_\infty$ structure
constructed in~\cite{Erler:2013xta} with a choice of $A_\delta (t)$.
It was also confirmed in~\cite{Erler:2015rra}
that the nonvanishing $F_{\delta t} (t)$ is annihilated by $D_\eta (t)$,
which is in accord with the general discussion.

\subsection{The equations of motion}

We are now ready to derive
the equations of motion from the action~\eqref{complete-action}.
We first show that the variation
$\delta \, \langle \, A_t (t), Q A_\eta (t) +( F (t) \Psi \, )^2 \, \rangle$
is a total derivative with respect to $t$.
The variation consists of three terms:
\begin{equation}
\begin{split}
& \delta \, \langle \, A_t (t), Q A_\eta (t) +( F (t) \Psi \, )^2 \, \rangle \\
& = \langle \, \delta A_t (t), Q A_\eta (t) +( F (t) \Psi \, )^2 \, \rangle
+\langle \, A_t (t), Q \, \delta A_\eta (t) \, \rangle
+\langle \, [ \, A_t (t), F (t) \Psi \, ], \delta F (t) \Psi \, \rangle \,.
\label{integrand-variation}
\end{split}
\end{equation}
The first term on the right-hand side of~\eqref{integrand-variation}
can be transformed as follows:
\begin{equation}
\begin{split}
& \langle \, \delta A_t (t), Q A_\eta (t) +( F (t) \Psi \, )^2 \, \rangle \\
& = \langle \, \partial_t A_\delta (t), Q A_\eta (t) +( F (t) \Psi \, )^2 \, \rangle
+\langle \, [ \, A_\delta (t), A_t (t) \, ], Q A_\eta (t) +( F (t) \Psi \, )^2 \, \rangle \\
& \quad~ +\langle \, F_{\delta t} (t), Q A_\eta (t) +( F (t) \Psi \, )^2 \, \rangle \\
& = \langle \, \partial_t A_\delta (t), Q A_\eta (t) +( F (t) \Psi \, )^2 \, \rangle
+\langle \, [ \, A_\delta (t), A_t (t) \, ], Q A_\eta (t) +( F (t) \Psi \, )^2 \, \rangle \,,
\end{split}
\end{equation}
where we used
\begin{equation}
\langle \, F_{\delta t} (t), Q A_\eta (t) +( F (t) \Psi \, )^2 \, \rangle
= \langle \, \{ D_\eta (t), f (t) \xi_0 \, \} \, F_{\delta t} (t), Q A_\eta (t) +( F (t) \Psi \, )^2 \, \rangle
= 0
\end{equation}
because $F_{\delta t} (t)$ and $Q A_\eta (t) +( F (t) \Psi \, )^2$
are annihilated by $D_\eta (t)$.
The second term on the right-hand side of~\eqref{integrand-variation}
can be transformed as follows:
\begin{equation}
\begin{split}
\langle \, A_t (t), Q \, \delta A_\eta (t) \, \rangle
& = \langle \, A_t (t), Q \, D_\eta (t) \, A_\delta (t) \, \rangle \\
& = \langle \, A_t (t), \{ Q, D_\eta (t) \} \, A_\delta (t) \, \rangle
-\langle \, A_t (t), D_\eta (t) \, Q A_\delta (t) \, \rangle \,.
\end{split}
\end{equation}
Using the identity
\begin{equation}
\{ Q, D_\eta (t) \} \, A = {}-[ \, Q A_\eta (t), A \, ] \,,
\label{Q-D_eta}
\end{equation}
we find
\begin{equation}
\begin{split}
\langle \, A_t (t), Q \, \delta A_\eta (t) \, \rangle
& = \langle \, A_t (t), [ \, A_\delta (t), Q A_\eta (t) \, ] \, \rangle
+\langle \, D_\eta (t) \, A_t (t), Q A_\delta (t) \, \rangle \\
& = \langle \, \partial_t A_\eta (t), Q A_\delta (t) \, \rangle
+\langle \, A_t (t), [ \, A_\delta (t), Q A_\eta (t) \, ] \, \rangle \\
& = \langle \, A_\delta (t), \partial_t Q A_\eta (t) \, \rangle
+\langle \, A_t (t), [ \, A_\delta (t), Q A_\eta (t) \, ] \, \rangle \,.
\end{split}
\end{equation}

To transform the third term on the right-hand side of~\eqref{integrand-variation},
let us calculate $\partial_t F (t) \Psi$ and $\delta F (t) \Psi$.
For $\partial_t F (t) \Psi$, we find
\begin{equation}
\partial_t F (t) \, \Psi
= [ \, \partial_t, F (t) \, ] \, \Psi
= {}-F (t) \, [ \, \partial_t, F^{-1} (t) \, ] \, F(t) \, \Psi
= {}-F(t) \, \Xi \, [ \, \partial_t, D_\eta (t) \, ] \, F(t) \, \Psi \,,
\end{equation}
where the actions of $[ \, \partial_t, F (t) \, ]$ and $[ \, \partial_t, F^{-1} (t) \, ]$
should be understood as
\begin{equation}
\begin{split}
[ \, \partial_t, F (t) \, ] \, A(t)
& = \partial_t F (t) \, A (t) -F (t) \, \partial_t A (t) \,, \\
[ \, \partial_t, F^{-1} (t) \, ] \, A(t)
& = \partial_t F^{-1} (t) \, A (t) -F^{-1} (t) \, \partial_t A (t) \,.
\end{split}
\end{equation}
We then use~\eqref{partial_t-D_eta} to obtain
\begin{equation}
\begin{split}
\partial_t F (t) \Psi
= F(t) \, \Xi \, \{ \, D_\eta (t) A_t (t), F(t) \Psi \}
& = F(t) \, \Xi \, D_\eta (t) \, [ \, A_t (t), F(t) \Psi \, ] \,.
\label{partial_t-F-Psi}
\end{split}
\end{equation}
For $\delta F (t) \Psi$, we find
\begin{equation}
\begin{split}
\delta F(t) \Psi
& = [ \, \delta, F(t) \, ] \, \Psi +F(t) \, \delta \Psi
= {}-F (t) \, [ \, \delta, F^{-1} (t) \, ] \, F (t) \, \Psi +F(t) \, \delta \Psi \\
& = {}-F(t) \, \Xi \, [ \, \delta, D_\eta (t) \, ] \, F(t) \Psi +F(t) \, \delta \Psi \,,
\end{split}
\end{equation}
where the actions of $[ \, \delta, F (t) \, ]$ and $[ \, \delta, F^{-1} (t) \, ]$
should be understood as
\begin{equation}
\begin{split}
[ \, \delta, F (t) \, ] \, A(t)
& = \delta F (t) \, A (t) -F (t) \, \delta A (t) \,, \\
[ \, \delta, F^{-1} (t) \, ] \, A(t)
& = \delta F^{-1} (t) \, A (t) -F^{-1} (t) \, \delta A (t) \,.
\end{split}
\end{equation}
We then use~\eqref{delta-D_eta} to obtain
\begin{equation}
\begin{split}
\delta F(t) \Psi
& = F(t) \, \Xi \, \{ \, D_\eta (t) A_\delta (t), F(t) \Psi \} +F(t) \, \eta \, \Xi \, \delta \Psi \\
& = F(t) \, \Xi \, D_\eta (t) \, [ \, A_\delta (t), F(t) \Psi \, ] +D_\eta (t) F(t) \, \Xi \, \delta \Psi \\
& = [ \, A_\delta (t), F(t) \Psi \, ]
+D_\eta (t) \, F(t) \, \Xi \, ( \, \delta \Psi -[ \, A_\delta (t), F(t) \Psi \, ] \, ) \,,
\end{split}
\end{equation}
where we also used~\eqref{D_eta-F} and~\eqref{D_eta-F-Xi}.
The third term on the right-hand side of~\eqref{integrand-variation}
can now be transformed as follows:
\begin{align}
& \langle \, [ \, A_t (t), F (t) \Psi \, ], \delta F (t) \Psi \, \rangle \nonumber \\
& = \langle \, [ \, A_t (t), F (t) \Psi \, ], [ \, A_\delta (t), F(t) \Psi \, ] \, \rangle
+\langle \, [ \, A_t (t), F (t) \Psi \, ],
D_\eta (t) \, F(t) \, \Xi \, ( \, \delta \Psi -[ \, A_\delta (t), F(t) \Psi \, ] \, ) \, \rangle \nonumber \\
& = \langle \, A_t (t), [ \, A_\delta (t), ( F(t) \Psi )^2 \, ] \, \rangle
+\langle \, D_\eta (t) \, [ \, A_t (t), F (t) \Psi \, ],
F(t) \, \Xi \, ( \, \delta \Psi -[ \, A_\delta (t), F(t) \Psi \, ] \, ) \, \rangle \,.
\end{align}
Note that the structure of the second term on the right-hand side
of the last line is similar to that of $\partial_t F (t) \Psi$ in~\eqref{partial_t-F-Psi}.
In fact, the operator $F(t) \, \Xi$ is BPZ even:
\begin{equation}
\langle \, F(t) \, \Xi \, A, B \, \rangle
= (-1)^A \, \langle \, A, F(t) \, \Xi \, B \, \rangle \,.
\end{equation}
This can be shown using
\begin{equation}
F (t) = \frac{1}{1-\Xi \, ( \, \eta -D_\eta (t) \, )}
\end{equation}
as follows:
\begin{equation}
\begin{split}
& \langle \, F(t) \, \Xi \, A, B \, \rangle
= \sum_{n=0}^\infty \, \langle \, 
( \, \Xi \, ( \, \eta -D_\eta (t) \, ) \, )^n \, \Xi \, A, B \, \rangle \\
& = \sum_{n=0}^\infty \, (-1)^A \, \langle \, A,
\Xi \, ( \, ( \, \eta -D_\eta (t) \, ) \, \Xi \, )^n B \, \rangle
= (-1)^A \, \langle \, A, F(t) \, \Xi \, B \, \rangle \,.
\end{split}
\end{equation}
We thus find
\begin{equation}
\begin{split}
& \langle \, [ \, A_t (t), F (t) \Psi \, ], \delta F (t) \Psi \, \rangle \\
& = \langle \, A_t (t), [ \, A_\delta (t), ( F(t) \Psi )^2 \, ] \, \rangle
+\langle \, F(t) \, \Xi \, D_\eta (t) \, [ \, A_t (t), F (t) \Psi \, ],
\delta \Psi -[ \, A_\delta (t), F(t) \Psi \, ] \, \rangle \\
& = \langle \, A_t (t), [ \, A_\delta (t), ( F(t) \Psi )^2 \, ] \, \rangle
+\langle \, \partial_t F(t) \Psi,
\delta \Psi -[ \, A_\delta (t), F(t) \Psi \, ] \, \rangle \\
& = \langle \, A_t (t), [ \, A_\delta (t), ( F(t) \Psi )^2 \, ] \, \rangle
+\langle \, A_\delta (t), \partial_t \, ( \, F (t) \Psi \, )^2 \, \rangle
-\langle \, \delta \Psi, \partial_t F(t) \Psi \, \rangle \,.
\end{split}
\end{equation}
The sum of the three terms on the right-hand side of~\eqref{integrand-variation} is then
\begin{equation}
\begin{split}
& \delta \, \langle \, A_t (t), Q A_\eta (t) +( F (t) \Psi \, )^2 \, \rangle \\
& = \langle \, \partial_t A_\delta (t), Q A_\eta (t) +( F (t) \Psi \, )^2 \, \rangle
+\langle \, [ \, A_\delta (t), A_t (t) \, ], Q A_\eta (t) +( F (t) \Psi \, )^2 \, \rangle \\
& \quad~ +\langle \, A_\delta (t), \partial_t Q A_\eta (t) \, \rangle
+\langle \, A_t (t), [ \, A_\delta (t), Q A_\eta (t) \, ] \, \rangle \\
& \quad~ +\langle \, A_t (t), [ \, A_\delta (t), ( F(t) \Psi \, )^2 \, ] \, \rangle
+\langle \, A_\delta (t), \partial_t \, ( \, F (t) \Psi \, )^2 \, \rangle
-\langle \, \delta \Psi, \partial_t F(t) \Psi \, \rangle \\
& = \partial_t \, \langle \, A_\delta (t), Q A_\eta (t) +( F (t) \Psi \, )^2 \, \rangle
-\partial_t \, \langle \, \delta \Psi, F(t) \Psi \, \rangle \,,
\end{split}
\end{equation}
where we used
\begin{equation}
\begin{split}
& \langle \, [ \, A_\delta (t), A_t (t) \, ], Q A_\eta (t) +( F (t) \Psi \, )^2 \, \rangle \\
& +\langle \, A_t (t), [ \, A_\delta (t), Q A_\eta (t) \, ] \, \rangle
+\langle \, A_t (t), [ \, A_\delta (t), ( F(t) \Psi )^2 \, ] \, \rangle = 0 \,.
\end{split}
\end{equation}
The variation $\delta \, \langle \, A_t (t), Q A_\eta (t) +( F (t) \Psi \, )^2 \, \rangle$
is a total derivative with respect to $t$ so that the $t$ dependence is topological.
This shows that the action is a functional of $\Phi$ and $\Psi$,
where $\Phi$ is the value of $\Phi (t)$ at $t=1$.
The variation of the action $\delta S$ is thus
\begin{equation}
\delta S = {}-\langle \, A_\delta, Q A_\eta +( F \Psi \, )^2 \, \rangle
+\langle \, \delta \Psi, F \Psi \, \rangle
-\llangle \, \delta \Psi, Y Q \Psi \, \rrangle \,.
\label{alternative-delta-S}
\end{equation}
The second term on the right-hand side can be transformed as
\begin{equation}
\begin{split}
& \langle \, \delta \Psi, F \Psi \, \rangle
= \langle \, \eta \, \xi_0 \, X Y \delta \Psi, F \Psi \, \rangle
= {}-\llangle \, X Y \delta \Psi, \eta F \Psi \, \rrangle
= {}-\llangle \, Y \delta \Psi, X \eta F \Psi \, \rrangle \\
& = {}-\llangle \, Y \delta \Psi, X Y X \eta F \Psi \, \rrangle
= {}-\llangle \, X Y \delta \Psi, Y X \eta F \Psi \, \rrangle
= {}-\llangle \, \delta \Psi, Y X \eta F \Psi \, \rrangle \,,
\end{split}
\end{equation}
and the final form of $\delta S$ is
\begin{equation}
\delta S = {}-\langle \, A_\delta, Q A_\eta +( F \Psi \, )^2 \, \rangle
-\llangle \, \delta \Psi, Y ( \, Q \Psi +X \eta F \Psi \, ) \, \rrangle \,.
\end{equation}
Therefore, the equations of motion are given by
\begin{subequations}
\label{equations-of-motion}
\begin{align}
Q A_\eta +( F \Psi \, )^2 & = 0 \,,
\label{NS-equation} \\
Q \Psi +X \eta F \Psi & = 0 \,.
\label{Ramond-equation}
\end{align}
\end{subequations}
Note that the second term on the left-hand side
of the equation of motion derived from $\delta \Psi$
is multiplied by $X \eta$.
The factor $\eta$ ensures that this term is in the small Hilbert space:
\begin{equation}
\eta X \eta F \Psi = 0 \,.
\end{equation}
The factor $X$ ensures that this term is in the restricted space:
\begin{equation}
X Y X \eta F \Psi = X \eta F \Psi
\end{equation}
because $X Y X = X$.
As we mentioned in the introduction,
this is the structure that we anticipated from the approach by Sen in~\cite{Sen:2015hha}.

\subsection{The gauge invariance}

Our remaining task is to derive
the gauge transformations~\eqref{gauge-transformations}.
When we set $\Psi = 0$, the action~\eqref{complete-action}
coincides with the WZW-like action $S_{\rm WZW}$
shown in~\eqref{S_WZW-1} or in~\eqref{S_WZW-2},
and it is invariant under the gauge transformations,
\begin{equation}
\delta_\Lambda^{(NS)} S_{\rm WZW} = 0 \,, \qquad
\delta_\Omega^{(NS)} S_{\rm WZW} = 0 \,,
\end{equation}
with $\delta_\Lambda^{(NS)} \Phi$ and $\delta_\Omega^{(NS)} \Phi$ given by
\begin{equation}
A_{\delta_\Lambda^{(NS)}} = Q \Lambda \,, \qquad
A_{\delta_\Omega^{(NS)}} = D_\eta \Omega \,,
\end{equation}
where $A_{\delta_\Lambda^{(NS)}}$ is $A_\delta$
with $\delta \Phi = \delta_\Lambda^{(NS)} \Phi$
and $A_{\delta_\Omega^{(NS)}}$ is $A_\delta$
with $\delta \Phi = \delta_\Omega^{(NS)} \Phi$.
Let us calculate the variations of $S$ in~\eqref{complete-action}
under $\delta_\Lambda^{(NS)} \Phi$ and $\delta_\Omega^{(NS)} \Phi$.

First, the variation $\delta_\Omega^{(NS)} S$ is given by
\begin{equation}
\delta_\Omega^{(NS)} S = {}-\langle \, D_\eta \Omega, Q A_\eta +( F \Psi )^2 \, \rangle
= {}-\langle \, \Omega, D_\eta \, ( \, Q A_\eta +( F \Psi )^2 \, ) \, \rangle = 0
\end{equation}
because $Q A_\eta$ and $F \Psi$ are annihilated by $D_\eta$.
Therefore, there are no corrections
to $\delta_\Omega^{(NS)}$ from the inclusion of the Ramond sector,
and we find
\begin{equation}
\delta_\Omega S = 0
\end{equation}
with
\begin{equation}
A_{\delta_\Omega} = D_\eta \Omega \,, \qquad \delta_\Omega \Psi = 0 \,,
\end{equation}
where $A_{\delta_\Omega}$ is $A_\delta$ with $\delta \Phi = \delta_\Omega \Phi$.

Let us next calculate the variation $\delta_\Lambda^{(NS)} S$:
\begin{equation}
\delta_\Lambda^{(NS)} S = {}-\langle \, Q \Lambda, Q A_\eta +( F \Psi )^2 \, \rangle
= {}-\langle \, \Lambda, Q \, ( \, Q A_\eta +( F \Psi )^2 \, ) \, \rangle
= \langle \, \{ F \Psi, \Lambda \}, Q F \Psi \, \rangle \,.
\end{equation}
The string field $Q F \Psi$ is given by
\begin{equation}
\begin{split}
Q F \Psi & = [ \, Q, F \, ] \, \Psi +F Q \Psi
= {}-F \, [ \, Q, F^{-1} \, ] \, F \, \Psi +F Q \Psi \\
& = {}-F \, [ \, Q, 1-\Xi \eta +\Xi D_\eta \, ] \, F \Psi +F Q \Psi \\
& = F \, ( \, X \eta -\{ Q, \Xi \} D_\eta +\Xi \, \{ Q, D_\eta \} \, ) \, F \Psi +F Q \Psi \,.
\end{split}
\end{equation}
Using $D_\eta F \Psi = 0$ and the identity~\eqref{Q-D_eta}, we have
\begin{equation}
Q F \Psi = F ( Q \Psi +X \eta F \Psi ) -F \Xi \, [ \, Q A_\eta, F \Psi \, ] \,.
\end{equation}
Note that $Q \Psi +X \eta F \Psi$ in the equation of motion~\eqref{Ramond-equation}
appeared in the first term on the right-hand side.
Since 
$[ \, ( F \Psi )^2, F \Psi \, ]=0\,$,
we can also make
$Q A_\eta +( F \Psi \, )^2$ in the equation of motion~\eqref{NS-equation}
appear in the second term on the right-hand side:
\begin{equation}
Q F \Psi = F ( Q \Psi +X \eta F \Psi ) -F \Xi \, [ \, Q A_\eta +( F \Psi )^2, F \Psi \, ] \,.
\label{Q-F-Psi}
\end{equation}
We can further transform $Q F \Psi$ as follows:
\begin{equation}
\begin{split}
Q F \Psi & = F \eta \Xi \, ( Q \Psi +X \eta F \Psi ) -F \Xi \, [ \, Q A_\eta +( F \Psi )^2, F \Psi \, ] \\
& = D_\eta F \Xi \, ( Q \Psi +X \eta F \Psi ) +F \Xi \, [ \, F \Psi, Q A_\eta +( F \Psi )^2 \, ] \,.
\end{split}
\end{equation}
Since $D_\eta$, $F \Xi$, and the graded commutator with $F \Psi$
are BPZ odd, BPZ even, and BPZ odd, respectively,
any BPZ inner product with $Q F \Psi$ can be brought
to a sum of an inner product with $Q \Psi +X \eta F \Psi$
and an inner product with $Q A_\eta +( F \Psi )^2$.
This allows the nonvanishing variation $\delta_\Lambda^{(NS)} S$
to be canceled by correcting the gauge transformations.
We find
\begin{equation}
\begin{split}
\delta_\Lambda^{(NS)} S
& = \langle \, \{ F\Psi, \Lambda \}, D_\eta F \Xi \, ( \, Q \Psi +X \eta F \Psi \, ) \, \rangle
+\langle \, \{ F \Psi, \Lambda \}, F \Xi \, [ \, F \Psi, Q A_\eta +( F \Psi )^2 \, ] \, \rangle \\
& = \langle \, F \Xi D_\eta \, \{ F \Psi, \Lambda \}, Q \Psi +X \eta F \Psi \, \rangle
+ \langle \, \{ F \Psi, F \Xi \, \{ F \Psi, \Lambda \} \}, Q A_\eta +( F \Psi )^2 \, \rangle \,.
\end{split}
\end{equation}
Since
\begin{equation}
\langle \, Q \Lambda +\{ F \Psi, F\Xi \, \{ F\Psi, \Lambda \} \}, Q A_\eta +(F\Psi)^2 \, \rangle
+\langle F\Xi D_\eta\{F\Psi,\Lambda\}, Q\Psi+X\eta F\Psi \, \rangle = 0 \,,
\end{equation}
we conclude that
\begin{equation}
\delta_\Lambda S = 0
\end{equation}
with
\begin{equation}
A_{\delta_\Lambda} = Q \Lambda +\{ F \Psi, F \Xi \, \{ F\Psi, \Lambda \} \} \,, \qquad
\delta_\Lambda \Psi = X \eta \, F \Xi D_\eta \, \{ F\Psi, \Lambda \} \,,
\end{equation}
where $A_{\delta_\Lambda}$ is $A_\delta$ with $\delta \Phi = \delta_\Lambda \Phi$.

Finally, let us derive the correction to the gauge transformation
\begin{equation}
\delta_\lambda^{(0)} \Phi = 0 \,, \qquad
\delta_\lambda^{(0)} \Psi = Q \lambda \,.
\end{equation}
We use the form of the variation of $S$ in~\eqref{alternative-delta-S} to find
\begin{equation}
\delta_\lambda^{(0)} S
= \langle \, Q \lambda, F \Psi \, \rangle
-\llangle \, Q \lambda, Y Q \Psi \, \rrangle
= {}-\langle \, \lambda, Q F \Psi \, \rangle \,.
\end{equation}
This takes the form of an inner product with $Q F \Psi$
so that it can be canceled by correcting the gauge transformation. We find
\begin{equation}
\begin{split}
\delta_\lambda^{(0)} S & = {}-\langle \, \lambda, D_\eta F \Xi \, ( \, Q \Psi +X \eta F \Psi \, ) \, \rangle
-\langle \, \lambda, F \Xi \, [ \, F \Psi, Q A_\eta +(F\Psi)^2 \, ] \, \rangle \\
& = {}-\langle \, F \Xi D_\eta \lambda, Q\Psi +X \eta F\Psi \, \rangle
-\langle \, F \Xi \lambda, [ \, F \Psi, Q A_\eta +(F\Psi)^2 \,] \, \rangle \\
& = {}-\langle \, F \Xi D_\eta \lambda, Q\Psi + X \eta F\Psi \, \rangle
-\langle \, \{ F \Psi, F \Xi \lambda \}, Q A_\eta +(F\Psi)^2 \, \rangle \,.
\end{split}
\end{equation}
We thus conclude that
\begin{equation}
\delta_\lambda S = 0
\end{equation}
with
\begin{equation}
A_{\delta_\lambda} = {}-\{F\Psi,F\Xi\lambda\} \,, \qquad
\delta_\lambda \Psi = Q \lambda -X \eta F\Xi D_\eta \lambda \,,
\end{equation}
where $A_{\delta_\lambda}$ is $A_\delta$ with $\delta \Phi = \delta_\lambda \Phi$.

In Section~\ref{cubic-quartic-section},
we chose the form of the gauge transformations
with the gauge parameter $\lambda$
such that $\lambda$ appears in the combination $\Xi \lambda$ except for $Q \lambda$.
While $\lambda$ in $A_{\delta_\lambda}$ appears in the combination $\Xi \lambda$,
it is not the case for the term $X \eta F\Xi D_\eta \lambda$ in~$\delta_\lambda \Psi$.
Using~\eqref{D_eta-F-Xi} and $\eta \lambda = 0$,
we can bring $\delta_\lambda \Psi$
to this form in the following way:
\begin{equation}
\delta_\lambda \Psi
= Q \lambda -X \eta \, ( \, 1 -D_\eta F\Xi \, ) \, \lambda
= Q \lambda +X \eta D_\eta F\Xi \lambda \,.
\label{delta_lambda-Psi-2}
\end{equation}
Since $X \eta D_\eta F \Xi \lambda = -X \eta \, \{ A_\eta, F \Xi \lambda \} \,$,
we see that this reproduces
$\delta_\lambda^{(1)} \Psi$ and $\delta_\lambda^{(2)} \Psi$
in Section~\ref{cubic-quartic-section}.
Furthermore, the gauge transformation $\delta_\lambda \Psi$ in~\eqref{delta_lambda-Psi-2}
can be brought to the form
\begin{equation}
\delta_\lambda \Psi
= Q \lambda +X \eta F \lambda
\label{delta_lambda-Psi-3}
\end{equation}
because $F \lambda = F \eta \Xi \lambda = D_\eta F \Xi \lambda \,$.
Note that the right-hand side of~\eqref{delta_lambda-Psi-3}
has the same structure as the equation of motion
$Q \Psi +X \eta F \Psi = 0$
with $\Psi$ replaced by $\lambda$.
We expect that this structure will play a role
in the Batalin-Vilkovisky quantization.

\section{Relation to the Berkovits formulation}\label{Berkovits-section}
\setcounter{equation}{0}

As we mentioned in the introduction, the equations of motion 
of open superstring field theory
including the Ramond sector were constructed by Berkovits in~\cite{Berkovits:2001im}.
In this section we investigate the relation between the equations of motion in~\cite{Berkovits:2001im}
and ours.

The equations of motion in~\cite{Berkovits:2001im} are given by
\begin{equation}
 \eta \, ( \, e^{-\Phi} Q \, e^\Phi \, ) +( \, \eta \psi^B \, )^2 =\ 0 \,, \qquad
Q \, ( \, e^\Phi \, ( \, \eta \psi^B \, ) \, e^{-\Phi} \, ) =\ 0 \,, 
\end{equation}
where $\Phi$ is the string field in the NS sector
and $\psi^B$ is the string field in the Ramond sector.
Both string fields are in the large Hilbert space.
Let us discuss the relation between our string field $\Psi$ and the string field $\psi^B$.
Since $\psi^B$ is in the large Hilbert space,  it is convenient to uplift our string field
to the large Hilbert space as well.
We introduce the string field $\psi$ in the large Hilbert space by
\begin{equation}
\Psi = \eta \psi \,.
\end{equation}
The condition that $\Psi$ is in the restricted space is translated into
\begin{equation}
X Y \eta \psi = \eta \psi \,.
\label{psi-constraint}
\end{equation}
The equations motion in terms of $\Phi$ and $\psi$ are
\begin{subequations}\label{psi}
\begin{align}
Q A_\eta +( F \eta \psi \, )^2 & = 0 \,,
\label{psi-NS} \\
Q \eta \psi +X \eta F \eta \psi & = 0 \,.
\label{psi-Ramond}
\end{align}
\end{subequations}

In order to find a relation between $\psi$ and $\psi^B$,
it is convenient to introduce $\tilde{\psi}^B$ defined by
\begin{equation}
\widetilde{\psi}^B = i  \, e^\Phi \psi^B e^{-\Phi} \,.
\end{equation}
Then the equations of motion in terms of $\Phi$ and $\tilde{\psi}^B$
are given by
\begin{subequations}\label{psi-tilde^B}
\begin{align}
Q A_\eta +( D_\eta \widetilde{\psi}^B )^2 = 0 \,,
\label{psi-tilde^B-NS} \\
Q D_\eta \widetilde{\psi}^B = 0 \,.
\label{psi-tilde^B-Ramond}
\end{align}
\end{subequations}
Since
\begin{equation}
F \eta \psi = D_\eta F \psi \,,
\end{equation}
where we used~\eqref{D_eta-F}, 
the equations of motion~\eqref{psi-NS} and~\eqref{psi-tilde^B-NS} in the NS sector coincide
under the field redefinition
\begin{equation}
\widetilde{\psi}^B = F \psi \,.
\label{field-redefinition}
\end{equation}

Let us next consider the equation of motion in the Ramond sector.
When $\Phi$ and $\psi$ satisfy the equations of motion~\eqref{psi},
the string fields $\Phi$ and $\widetilde{\psi}^B$
mapped by the field redefinition~\eqref{field-redefinition}
satisfy the equation of motion~\eqref{psi-tilde^B-Ramond} because
\begin{equation}
\begin{split}
Q D_\eta \widetilde{\psi}^B
= Q F \eta \psi
= F ( Q \eta \psi +X \eta F \eta \psi ) -F \Xi \, [ \, Q A_\eta +( F \eta \psi )^2, F \eta \psi \, ] \,,
\label{R-eom-mapping}
\end{split}
\end{equation}
where we used~\eqref{Q-F-Psi}.
On the other hand, we can transform~\eqref{R-eom-mapping} as
\begin{equation}
\begin{split}
Q \eta \psi +X \eta F \eta \psi
& = F^{-1} Q F \eta \psi
+\Xi \, [ \, Q A_\eta +( F \eta \psi )^2, F \eta \psi \, ] \\
& = F^{-1} Q D_\eta \widetilde{\psi}^B
+\Xi \, [ \, Q A_\eta +( D_\eta \widetilde{\psi}^B )^2, D_\eta \widetilde{\psi}^B \, ] \,,
\end{split}
\end{equation}
so the string fields $\Phi$ and $\psi$ satisfy the equation of motion~\eqref{psi-Ramond}
when $\Phi$ and $\widetilde{\psi}^B$ satisfy
the equations of motion~\eqref{psi-tilde^B}.
We thus conclude that the two sets of the equations of motion~\eqref{psi-tilde^B}
and~\eqref{psi} are equivalent under the field redefinition~\eqref{field-redefinition}.

Finally, let us see how the condition~\eqref{psi-constraint}
that $\eta \psi$ is in the restricted space
is mapped by the field redefinition~\eqref{field-redefinition}.
Since
\begin{equation}
\eta F^{-1} = \eta \Xi D_\eta \,,
\end{equation}
the condition~\eqref{psi-constraint} is translated into
the following condition on~$\widetilde{\psi}^B$:
\begin{equation}
X Y \eta \Xi D_\eta \widetilde{\psi}^B = \eta \Xi D_\eta \widetilde{\psi}^B \,.
\end{equation}
We can also translate it into the condition on $\psi^B$ as
\begin{equation}
X Y \eta \Xi \, ( \, e^\Phi \, ( \, \eta \psi^B \, ) \, e^{-\Phi} \, )
= \eta \Xi \, ( \, e^\Phi \, ( \, \eta \psi^B \, ) \, e^{-\Phi} \, ) \,.
\end{equation}
Both forms of the constraint are highly nontrivial
since they are conditions on nonlinear combinations of string fields 
involving not only the string field in the Ramond sector
but also the string field in the NS sector. 
This suggests that our choice of the string field in the Ramond sector,
$\Psi$ or $\psi$, is canonical
in constructing an action,
and no field redefinitions in the Ramond sector seem to be allowed.

\section{Conclusions and discussion}\label{conclusions-discussion}
\setcounter{equation}{0}

In this paper we constructed the action~\eqref{complete-action}
for open superstring field theory.
It includes both the NS sector and the Ramond sector,
and it is invariant under the gauge transformations given
by~\eqref{gauge-transformations}.
This is the first construction of a complete action for superstring field theory
in a covariant form.
The gauge invariance ensures the decoupling of unphysical states,
and we believe that correct scattering amplitudes at the tree level
will be reproduced~\cite{amplitudes}.\footnote{
See~\cite{Konopka:2015tta} for a mathematical discussion
on the S-matrix of superstring field theory.
}

We use the large Hilbert space for the NS sector
and the small Hilbert space for the Ramond sector.
Let us first discuss the possibility of formulations
within the framework of the small Hilbert space.
As we have already mentioned, our action can also be interpreted
as the action for the string fields in the small Hilbert space
both for the NS sector and the Ramond sector.
We only need to parameterize $A_\eta (t)$ and $A_t (t)$
satisfying the relations~\eqref{fundamental-relations}
in terms of a string field in the small Hilbert space.
We can use the partial gauge fixing in~\cite{Iimori:2013kha}
or use the string field in~\cite{Erler:2013xta} for the action
with the $A_\infty$ structure, as demonstrated in~\cite{Erler:2015rra}.
While the resulting theory is described in terms of
string fields in the small Hilbert space,
the structure of the large Hilbert space is used
in an essential way in these formulations.
In our context, it is manifested in the aspect
that we need to use the operator $\Xi$.
However, we do not foresee any fundamental obstructions
in constructing a gauge-invariant action
within the framework of the $\beta \gamma$ ghosts
by extending the approach in an upcoming paper~\cite{Ohmori-Okawa} based on the covering
of the supermoduli space of super-Riemann surfaces
to the Ramond sector.
The reason we use the large Hilbert space for the NS sector
is to have a closed-form expression for the action,
and it would be an important problem to construct an action
in a closed form based on the $\beta \gamma$ ghosts.

Let us next discuss the possibility of formulations
based on the large Hilbert space.
As we did in Section~\ref{Berkovits-section},
it is straightforward to uplift the string field in the Ramond sector
to the large Hilbert space, but the structure of the small Hilbert space is crucially used
in the characterization of the space of string fields in the Ramond sector
in terms of the operator $X$.
It would be more flexible if we could
characterize it in terms of $\xi (z)$, $\eta (z)$, and $\phi (z)$.
For example, in the approach by Sen~\cite{Sen:2015hha}, the zero mode $X_0$
of the picture-changing operator is used for the propagator in the Ramond sector,
and this seems to suggest a possibility of characterizing the space of string fields
in the Ramond sector in terms of $X_0$, as the information on degrees of freedom
should be reflected in the propagator.
If this is possible, we may be able to replace $\Xi$ with the zero mode $\xi_0$,
as the origin of the operator $\Xi$ is the relation $X = \{ Q, \Xi \}$
and $X_0$ can be written as $X_0 = \{ Q, \xi_0 \}$.
Use of the large Hilbert space obscures the relation to the supermoduli space
of super-Riemann surfaces at the moment,
and we had to use the framework of the $\beta \gamma$ ghosts
in describing the space of string fields in the Ramond sector.
We hope to have formulations of superstring field theory
where the large Hilbert space and the supermoduli space of super-Riemann surfaces
are integrated in a fundamental fashion.

Now that we have the complete action~\eqref{complete-action} for open superstring field theory,
we can address interesting questions.
First of all, we are now at a starting point for quantizing open superstring field theory.
One important question that
we can address by quantizing open superstring field theory would be 
whether we can describe closed strings in terms of open string fields
or whether we need closed string fields as independent degrees of freedom.
The first step is the construction of a classical master action
in the Batalin-Vilkovisky formalism~\cite{Batalin:1981jr, Batalin:1984jr} for quantization.
As we mentioned before, the action in the NS sector
can be described by multi-string products
satisfying the $A_\infty$ relations~\cite{Erler:2013xta, Erler:2015rra, Erler:2015uba},
and the Batalin-Vilkovisky quantization is straightforward.
In addition, as we commented at the end of Section~\ref{complete-action-section},
the equation of motion~\eqref{Ramond-equation} in the Ramond sector
and the gauge transformation $\delta_\lambda \Psi$ in~\eqref{delta_lambda-Psi-3}
share the same structure,
which is promising for the Batalin-Vilkovisky quantization.
It would also be a promising approach
to adapt the recent construction of the equations of motion
including the Ramond sector
in terms of multi-string products
satisfying the $A_\infty$ relations~\cite{Erler:2015lya}
so that string products in the Ramond sector
are consistent with the projection to the restricted space.

Another important question that we can discuss with our action
would be spacetime supersymmetry preserved by the D-brane.
It would again be helpful to see~\cite{Erler:2015lya}
for a recent discussion
on supersymmetry based on the equations of motion.\footnote{
See also~\cite{Kishimoto:2005bs} for a different approach. 
}
A more ambitious question would be to uncover
the supersymmetry spontaneously broken by the presence of the D-brane.

It would also be fascinating to extend the present formulation to closed superstring field theory. 
While the construction of a complete action for type II superstring field theory
seems challenging,
we hope that our construction can be extended
to heterotic string field theory~\cite{Okawa:2004ii,Berkovits:2004xh,Kunitomo:2013mqa,Kunitomo:2014hba}.\footnote{
After we submitted this paper to arXiv, master actions in the Batalin-Vilkovisky formalism
were constructed for heterotic string field theory and for type II superstring field theory
in~\cite{Sen:2015uaa}, where covariant kinetic terms are constructed
by introducing additional free fields.
}

Construction of a complete action for open superstring field theory
is not the end of the story.
It is just the beginning.
We hope that this will provide a useful approach
complementary to other directions such as the AdS/CFT correspondence
and help us unveil the nature of the nonperturbative theory underlying the perturbative superstring theory.

\bigskip
\noindent
{\bf \large Acknowledgments}

\medskip
We would like to thank Ted Erler and Ashoke Sen for helpful discussions.
We also thank the Center for Theoretical Physics,
College of Physical Science and Technology, Sichuan University
for hospitality during
the ``International Conference on String Field Theory and Related Aspects VII'',
where this work was initiated.
The work of Y.O. was supported in part
by a Grant-in-Aid for Scientific Research~(B) No.~25287049
and a Grant-in-Aid for Scientific Research~(C) No.~24540254
from the Japan Society for the Promotion of Science (JSPS).

\appendix

\section{The integration over the fermionic modulus}
\setcounter{equation}{0}
\label{appendix-X}

As we mentioned in the introduction, the operator $X$ given by
\begin{equation}
X = {}-\delta ( \beta_0 ) \, G_0 +\delta' ( \beta_0 ) \, b_0 \, ,
\label{PCO-appendix}
\end{equation}
which is used to characterize the restricted space
of string fields in the Ramond sector,
is related to the integration of the fermionic modulus
of propagator strips in the Ramond sector.
In this appendix we elaborate on this aspect and show that the expression~\eqref{PCO-appendix}
can be obtained from the expression
\begin{equation}
X = \int d \zeta \int d \tilde{\zeta} \, e^{\, \zeta G_0 -\tilde{\zeta} \, \beta_0}
\label{X-appendix}
\end{equation}
by carrying out the integration over the fermionic modulus $\zeta$.\footnote{
This appendix is based on the results for the NS sector
in an upcoming paper~\cite{Ohmori-Okawa}.
}

In~\cite{Witten:2012bh} the extended BRST transformation was introduced,
and the fermionic modulus $\zeta$ is mapped to the Grassmann-even variable $\tilde{\zeta}$
by the extended BRST transformation.
The extended BRST transformation acts in the same way
as the ordinary BRST transformation for operators in the boundary CFT,
and in particular it maps $\beta_0$ to $G_0$.
Therefore, the combination $\zeta G_0 -\tilde{\zeta} \, \beta_0$
in~\eqref{X-appendix} is obtained from $-\zeta \beta_0$
by the extended BRST transformation.

Let us carry out the integration over $\zeta$ in~\eqref{X-appendix}.
Using the commutation relations
\begin{equation}
[ \, G_0, \beta_0 \, ] = {}-2 \, b_0 \,, \qquad
[ \, \beta_0, b_0 \, ] = 0 \,, \qquad
\{ G_0, b_0 \} = 0
\end{equation}
and the Baker-Campbell-Hausdorff formula,  we find
\begin{equation}
e^{\, \zeta G_0 -\tilde{\zeta} \, \beta_0}
= e^{-\frac{1}{2} \, [ \, \zeta G_0,  \tilde{\zeta} \, \beta_0 \, ]} \,
e^{-\tilde{\zeta} \, \beta_0} e^{\, \zeta G_0} 
= e^{\tilde{\zeta} \, \zeta \, b_0} \,
e^{-\tilde{\zeta} \, \beta_0} e^{\, \zeta G_0} 
= e^{-\tilde{\zeta} \, \beta_0} \, ( 1 + \tilde{\zeta} \, \zeta \, b_0 +\zeta G_0 ) \,.
\end{equation}
By integrating over $\zeta$, the operator $X$ can be written as
\begin{equation}
X = - \int d \tilde{\zeta} \int d \zeta \,
e^{-\tilde{\zeta} \, \beta_0} \, ( 1 + \tilde{\zeta} \, \zeta \, b_0 +\zeta G_0 ) 
= \int d \tilde{\zeta} \,
e^{-\tilde{\zeta} \, \beta_0} \, ( \, -\tilde{\zeta} \, b_0 - G_0 ) \,,
\end{equation}
where we treated $d \zeta$ and $d \tilde{\zeta}$ as Grassmann-odd objects.
As emphasized in~\cite{Witten:2012bh},
the integral over the Grassmann-even variable $\tilde{\zeta}$
should not be considered as an ordinary integral,
and it should be regarded as an algebraic operation
similar to the integration over Grassmann-odd variables.
See~\cite{Witten:2012bh} for more details.
In this context, we define $\delta ( \beta_0 )$ and $\delta' ( \beta_0 )$ by
\begin{equation}
\delta ( \beta_0 ) = \int d \tilde{\zeta} \, e^{-\tilde{\zeta} \, \beta_0} \,, \qquad
\delta' ( \beta_0 ) = -\int d \tilde{\zeta} \, \tilde{\zeta} \, e^{-\tilde{\zeta} \, \beta_0} \,,
\end{equation}
and the operator $X$ is written as
\begin{equation}
X = \delta' ( \beta_0 ) \, b_0 - \delta ( \beta_0 ) \, G_0 \,.
\end{equation}
Note that $\delta ( \beta_0 )$ and $\delta' ( \beta_0 )$ are Grassmann-odd operators
because we treat $d \zeta$ and $d \tilde{\zeta}$ as Grassmann-odd objects.
We have thus obtained the expression~\eqref{PCO-appendix} for $X$.

\section{Properties of $\Xi$}
\setcounter{equation}{0}
\label{appendix-Xi}

In this appendix we first show that the anticommutator of $\eta$ and $\Xi$ is given by
\begin{equation}
\{ \eta, \Xi \} = 1
\end{equation}
for
\begin{equation}
\Xi = \Theta ( \beta_0 ) \,,
\end{equation}
where $\Theta$ is the Heaviside step function.
We begin with the identification~\cite{Verlinde:1987sd}
\begin{equation}
\Theta ( \beta (\sigma) ) = \xi (\sigma) \,,
\end{equation}
where
\begin{equation}
\beta (\sigma) = \sum_n \beta_n \, e^{i n \sigma} \,, \qquad
\xi (\sigma) = \sum_n \xi_n \, e^{i n \sigma} \,.
\end{equation}
We then separate $\beta (\sigma)$ as
\begin{equation}
\beta (\sigma) = \beta_0 +\widetilde{\beta} (\sigma) \,,
\end{equation}
where
\begin{equation}
[ \, \gamma_0, \beta_0 \, ] = 1 \,, \qquad
[ \, \gamma_0, \widetilde{\beta} (\sigma) \, ] = 0 \,,
\end{equation}
and we rewrite $\xi (\sigma)$ in the following way:
\begin{equation}
\xi (\sigma) = \Theta ( \beta_0 +\widetilde{\beta} (\sigma) )
= e^{\, \widetilde{\beta} (\sigma) \, \gamma_0} \, \Theta ( \beta_0 ) \,
e^{-\widetilde{\beta} (\sigma) \, \gamma_0} \,.
\end{equation}
We invert this relation to write $\Xi$ in terms of $\xi (\sigma)$ as follows:
\begin{equation}
\Xi = \Theta ( \beta_0 )
= e^{-\widetilde{\beta} (\sigma) \, \gamma_0} \, \xi (\sigma) \,
e^{\, \widetilde{\beta} (\sigma) \, \gamma_0} \,.
\label{mode-Xi}
\end{equation}
It follows from~\eqref{bosonization} that
\begin{equation}
[ \, \eta, \widetilde{\beta} (\sigma) \, ] 
= [ \, \eta, \beta (\sigma) -\beta_0 \, ] = 0 \,, \qquad
[ \, \eta, \gamma_0 \, ] = 0 \,,
\end{equation}
and we also use
\begin{equation}
\{ \eta, \xi (\sigma) \} = 1
\end{equation}
to find
\begin{equation}
\{ \eta, \Xi \} = e^{-\widetilde{\beta} (\sigma) \, \gamma_0} \, \{ \eta, \xi (\sigma) \} \,
e^{\, \widetilde{\beta} (\sigma) \, \gamma_0} = 1 \,.
\end{equation}

Let us next show that $\Xi$ is BPZ even based on the expression~\eqref{mode-Xi}.
We denote the BPZ conjugate of an operator $\mathcal{O}$ by $\mathcal{O}^\star$.
Consider the mode expansion of a primary field $\varphi (z)$ of weight $h$.
In general, the BPZ conjugate of the mode $\varphi_n$
with $[ \, L_0, \varphi_n \, ] = -n \, \varphi_n$ is given by
\begin{equation}
\varphi_n^\star = (-1)^{n+h} \, \varphi_{-n} \,.
\end{equation}
We therefore have
\begin{equation}
\xi_n^\star = (-1)^n \, \xi_{-n} \,, \qquad
\gamma_n^\star = (-1)^{n-\frac{1}{2}} \, \gamma_{-n} \,, \qquad
 \beta_n^\star = (-1)^{n+\frac{3}{2}} \, \beta_{-n} \,.
\end{equation} 
The right-hand side of~\eqref{mode-Xi} is actually independent of $\sigma$,
and it is convenient to set $\sigma = \pi / 2$.
Since
\begin{equation}
\widetilde{\beta} \, \Bigl( \frac{\pi}{2} \Bigr)^\star
=  (-1)^{\frac{3}{2}} \, \widetilde{\beta} \, \Bigl( \frac{\pi}{2} \Bigr) \,, \qquad
\gamma_0^\star = (-1)^{-\frac{1}{2}} \, \gamma_0 \,, \qquad
\xi \, \Bigl( \frac{\pi}{2} \Bigr)^\star = \xi \, \Bigl( \frac{\pi}{2} \Bigr) \,,
\end{equation}
we find
\begin{equation}
\begin{split}
\Xi^\star & = 
\Bigl( e^{\, \widetilde{\beta} \, (\pi/2) \, \gamma_0} \Bigr)^\star \,
\xi \, \Bigl( \frac{\pi}{2} \Bigr)^\star
\Bigl( e^{-\widetilde{\beta} (\pi/2) \, \gamma_0} \Bigr)^\star \\
& = e^{-\gamma_0\, \widetilde{\beta} (\pi/2)} \, 
\xi \, \Bigl( \frac{\pi}{2} \Bigr) \, e^{\gamma_0 \,\widetilde{\beta} (\pi/2)}
= \Xi \,.
\end{split}
\end{equation}

\end{document}